\documentclass[a4paper,fleqn,usenatbib]{mnras}

\usepackage[T1]{fontenc}
\usepackage{ae,aecompl}

\usepackage{graphicx}	
\usepackage{amsmath}	
\usepackage{amssymb}	
\usepackage{deluxetable}
\usepackage{pdflscape}
\usepackage{verbatim}

\newcommand{\chandra}{\textit{Chandra}}
\newcommand{\xmm}{\textit{XMM-Newton}}
\newcommand{\civ}{C~{\sc iv}}
\newcommand{\heii}{He~{\sc ii}}
\newcommand{\feii}{Fe~{\sc ii}}
\newcommand{\feiii}{Fe~{\sc iii}}
\newcommand{\aox}{$\alpha_{\rm ox}$}
\newcommand{\daox}{$\Delta\alpha_{\rm ox}$}
\newcommand{\lopt}{$L_{\rm 2500~\mathring{\rm{A}}}$}
\newcommand{\fopt}{$f_{\rm 2500~\mathring{\rm{A}}}$}
\newcommand{\fx}{$f_{\rm 2~keV}$}

\newcommand{\flux}{{erg~cm$^{-2}$~s$^{-1}$}}
\newcommand{\mflux}{{erg~cm$^{-2}$~s$^{-1}$~Hz$^{-1}$}}
\newcommand{\lum}{{erg~s$^{-1}$}}
\newcommand{\mlum}{{erg~s$^{-1}$~Hz$^{-1}$}}

\newcommand{\xray}{\hbox{X-ray}}
\newcommand{\gi}{$\Delta (g-i)$}
\newcommand{\nh}{$N_{\rm H}$}

\title[Full range of X-ray properties of WLQs]{Sensitive \textit{Chandra} coverage of a representative sample of weak-line quasars: revealing the full range of X-ray properties}

\author[Q. Ni et al.]{Q. Ni,$^{1,2,3}$\thanks{E-mail: qingling1001@gmail.com}
W. N. Brandt,$^{1,2,4}$
B. Luo,$^{5,6}$
G. P. Garmire,$^{7}$
P. B. Hall,$^{8}$
R. M. Plotkin,$^{9}$\newauthor
O. Shemmer,$^{10}$
J. D. Timlin III,$^{1,2}$
F. Vito,$^{11}$
J. Wu,$^{12}$
and W. Yi$^{1,13,14}$
\\
$^{1}$Department of Astronomy and Astrophysics, 525 Davey Lab, The Pennsylvania State University, University Park, PA 16802, USA\\
$^{2}$Institute for Gravitation and the Cosmos, The Pennsylvania State University, University Park, PA 16802, USA\\
$^{3}$Institute for Astronomy, University of Edinburgh, Royal Observatory, Edinburgh, EH9 3HJ, UK\\
$^{4}$Department of Physics, The Pennsylvania State University, University Park, PA 16802, USA\\
$^{5}$School of Astronomy and Space Science, Nanjing University, Nanjing, Jiangsu 210093, China\\
$^{6}$Key Laboratory of Modern Astronomy and Astrophysics (Nanjing University), Ministry of Education, Nanjing, Jiangsu 210093, China\\
$^{7}$Huntingdon Institute for X-ray Astronomy, LLC, 10677 Franks Road, Huntingdon, PA 16652, USA\\
$^{8}$Department of Physics \& Astronomy, York University, 4700 Keele Street, Toronto, ON M3J 1P3, Canada\\
$^{9}$Department of Physics, University of Nevada, 1664 N. Virginia St, Reno, Nevada, 89557, USA\\
$^{10}$Department of Physics, University of North Texas, Denton, TX 76203, USA\\
$^{11}$Scuola Normale Superiore, Piazza dei Cavalieri 7, 56126, Pisa, Italy\\
$^{12}$Department of Astronomy, Xiamen University, Xiamen, Fujian 361005, China\\
$^{13}$Yunnan Observatories, Kunming, 650216, China\\
$^{14}$Key Laboratory for the Structure and Evolution of Celestial Objects, Chinese Academy of Sciences, Kunming 650216, China
}

\date{Accepted XXX. Received YYY; in original form ZZZ}

\pubyear{2021}

\begin{document}
\label{firstpage}
\pagerange{\pageref{firstpage}--\pageref{lastpage}}
\maketitle

\begin{abstract}
We present deeper \chandra\ observations for weak-line quasars (WLQs) in a representative sample that previously had limited X-ray constraints, and perform \xray\ photometric analyses to reveal the full range of X-ray properties of WLQs.
Only 5 of the 32 WLQs included in this representative sample remain \xray\ undetected after these observations, and a stacking analysis shows that these 5 have an average X-ray weakness factor of $> 85$.
One of the WLQs in the sample that was known to have extreme X-ray variability, SDSS~J1539+3954, exhibited dramatic X-ray variability again: it changed from an X-ray normal state to an X-ray weak state within $\approx$ 3 months in the rest frame. This short timescale for an X-ray flux variation by a factor of $\gtrsim 9$ further supports the thick disk and outflow (TDO) model proposed to explain the X-ray and multiwavelength properties of WLQs.
The overall distribution of the \xray-to-optical properties of WLQs suggests that the TDO has an average covering factor of the X-ray emitting region of $\sim$ 0.5, and the column density of the TDO can range from $N_{\rm H}$ $\sim 10^{23-24}~{\rm cm}^{-2}$ to $N_{\rm H}$ $\gtrsim 10^{24}~{\rm cm}^{-2}$, which leads to different levels of absorption and Compton reflection (and/or scattering) among WLQs.

\end{abstract}




\section{Introduction}

Investigations of the type~1 quasar sub-class of weak-line quasars (WLQs)
have recently provided valuable insights into quasar structure and accretion physics.
While WLQs are luminous blue quasars which are believed to be viewed generally face-on according to the standard unification model (e.g. \citealt{Antonucci1993, Netzer2015}), they have weak or no high-ionization emission lines
\citep[e.g.][]{Fan1999,DS2009,Plotkin2010,Wu2012}, which marks how they differ from typical blue quasars. For example, their C\,{\sc iv} rest-frame equivalent widths (REWs)
are \hbox{$\lesssim 10$--15~\AA}, corresponding to $\gtrsim 3\sigma$ negative deviations
from the mean. Their C\,{\sc iv} lines also often show very
large blueshifts of \hbox{1000--10000~km~s$^{-1}$}. The majority of these WLQs are radio
quiet.

X-ray studies of WLQs have proved particularly revealing for understanding their nature.
A number of remarkable WLQ \xray\ properties have been identified:

\begin{enumerate}

\item
WLQs show a wide range of \xray\ luminosities compared to expectations
from their optical/UV luminosities or overall spectral energy distributions (SEDs), when considering, e.g. \aox\ and \daox\ (e.g. \citealt{Wu2011, Wu2012, Luo2015, Ni2018}).
\aox\ is the slope of
a nominal power law connecting the rest-frame 2500~\AA\ and 2~keV monochromatic
luminosities; i.e., $\alpha_{\rm ox}=0.3838 \log(L_{\rm 2~keV}/L_{2500~\mathring{\rm{A}}})$ (e.g. \citealt{Tananbaum1979, Strateva2005}).
\aox\ reflects the relative strength of a quasar's X-ray emission to its UV emission, indicating a strong connection between the corona and the disk.
This quantity is correlated with $L_{2500~\mathring{\rm{A}}}$. 
We also define 
$\Delta\alpha_{\rm ox}=\alpha_{\rm ox}({\rm Observed})-\alpha_{\rm ox}(L_{2500~\mathring{\rm{A}}})$, 
which quantifies the deviation of the observed \xray\ 
luminosity relative to that expected from the 
\aox-$L_{2500~\mathring{\rm{A}}}$ relation (e.g. \citealt{Gallagher2006}).\footnote{\daox\ is used to derive 
factors of \xray\ weakness ($f_{\rm weak}=403^{-\Delta\alpha_{\rm ox}}$) in the text.}
\daox\ can thus serve as an indicator of a quasar's X-ray emission strength compared to the expectation from its UV emission.

About half of the WLQ population shows weak \xray\ emission, often by
notable factors of 10 or more; this fraction is much higher than present 
among the general quasar population \citep[e.g.][]{Pu2020, Timlin2020}. Typical blue quasars with weak \xray\ emission are mostly 
broad absorption line (BAL) quasars with high levels of intrinsic \xray\ absorption (e.g. \citealt{Brandt2000, Gallagher2002, Gallagher2006, Fan2009}), but the blue WLQs targeted in \xray\ observations have been selected not to have BALs,
mini-BALs, or other strong rest-frame UV absorption, which makes the high X-ray weak fraction among WLQs intriguing.
However, owing to a high fraction of \xray\ non-detections among \xray\ weak WLQs, 
the \xray\ weakness distribution remains poorly defined for large values
of \xray\ weakness. The other half of the WLQ population generally shows nominal-strength \xray\ emission.

\item 
X-ray spectral analyses of the members of the WLQ population with nominal-strength
\xray\ emission, utilizing both individual-object and spectral-stacking approaches,
show notably steep power-law spectra typically with photon indices
of \hbox{$\Gamma=2.1$--2.3} \citep{Luo2015, Marlar2018}. Among the
general quasar population, such steep power-law spectra indicate accretion
onto the supermassive black hole (SMBH) with high Eddington ratio ($L/L_{\rm Edd}$;
e.g. \citealt{Shemmer2008,Brightman2013}), and thus it seems likely that WLQs generally have high $L/L_{\rm Edd}$.
This conclusion is also supported by the \civ\ REWs and blueshifts of WLQs (e.g. \citealt{Luo2015, Ni2018}), as well as rough single-epoch estimates of $L/L_{\rm Edd}$ $\approx 0.3$--1.3 (e.g. \citealt{Luo2015,Plotkin2015}).

\item
X-ray spectral analyses of the \xray\ weak members of the WLQ population
are more challenging, owing to limited numbers of detected counts in the
available observations. Stacking analyses of the \xray\ weak WLQs
generally show hard \xray\ spectra, on average, with effective power-law
photon indices of $\langle \Gamma \rangle\approx 1.2$--1.4 \citep{Luo2015, Ni2018}. 
These hard \xray\ spectra suggest high levels of intrinsic \xray\
absorption ($N_{\rm H}\gtrsim 10^{23}$~cm$^{-2}$), which 
is somewhat surprising given these quasars' type~1 nature and blue
optical/UV continua without BAL (or other UV absoprtion) features.
When such heavy \xray\ absorption is present, Compton reflection and/or scattering
may also play significant roles in shaping the \xray\ spectrum. 
The presence of heavy \xray\ absorption is further supported by an 
individual \chandra\ spectrum of SDSS~J1521+5202, an extraordinarily  
luminous WLQ \citep{Luo2015}, and studies of some local WLQ analogs (e.g. \citealt{Reeves2020}).

\item
The current limited multi-epoch \xray\ observations of WLQs suggest that
they show extreme large-amplitude \xray\ variability events
\citep[e.g.][]{Miniutti2012, Ni2020} more frequently
than the general radio-quiet quasar population of comparable luminosity
\citep[e.g.][]{Timlin2020a}. The observed large-amplitude \xray\ variability
events are associated with changes between X-ray normal-strength and X-ray weak states, and they
do not appear to have corresponding extreme optical/UV continuum or
emission-line variations. 

\end{enumerate}

\noindent
The above \xray\ and multiwavelength results for WLQs have been used to construct a
basic scenario that appears able to explain consistently their weak
high-ionization lines, their \xray\ properties, and some of 
their other multiwavelength properties \citep[e.g.][]{Lane2011, Wu2011, Wu2012,
Luo2015, Ni2018}. This scenario, depicted in Fig.~1 of \citet{Ni2018}, proposes that, owing to high $L/L_{\rm Edd}$ ($\gtrsim 0.3$), WLQs
have geometrically and optically thick inner accretion disks and associated
outflows \citep[e.g.][]{Jiang2014, Jiang2019, Wang2014, Dai2018}.
The thick disk and its outflow (hereafter, TDO), which lie near the central supermassive black hole, prevent ionizing EUV/\xray\
photons from reaching much of the (substantially equatorial) high-ionization
broad emission-line region (BLR), leading to the weak high-ionization lines. The
TDO is also responsible for the \xray\ weakness/absorption seen in about
half of WLQs; \xray\ weak WLQs arise for systems viewed at large inclinations,
so that our line of sight intercepts the TDO. The large-amplitude \xray\
variability events of some WLQs may arise from slight variations in the
thickness of the TDO (e.g. rotation of an inner TDO that is somewhat
azimuthally asymmetric) or internal motions of the TDO. 

Our primary approach to investigating the \xray\ properties of WLQs has
involved \chandra\ ``snapshot'' (typically \hbox{2--5~ks}) observations 
of luminous (typically $M_i\lesssim -27$), blue WLQs.\footnote{These ``snapshot'' observations were largely aimed at efficiently obtaining the minimum number of counts needed to estimate the \xray\ flux levels of these sources.}
While this approach has
been successful for gaining a broad overview of basic WLQ \xray\ properties,
it has provided only limited insight into the \xray\ properties of
individual WLQs, especially for the most interesting \xray\ weak members
of this class. We know that many of these objects must be remarkably \xray\
weak from \xray\ stacking analyses, having \daox\ of $-0.4$ to $-0.7$, 
corresponding to \xray\ weakness by factors of $\approx 11$--67
relative to a typical radio-quiet quasar. For comparison, our constraints
on individual \xray\ undetected objects typically can only show they have
$\Delta\alpha_{\rm ox}\lesssim -0.3$, corresponding to \xray\ weakness by a
factor of $\gtrsim 6$. We thus currently lack useful information on the
full distribution of \daox\ for WLQs, which should provide information about
the nature of the TDO. Moreover, the limited individual-object \xray\
constraints have hindered our ability to assess UV continuum and
emission-line properties that may correlate with \xray\ weakness, thereby
serving as useful tracers of this behavior. 

We have therefore performed additional, deeper (typically 5--23 ks) \chandra\ observations for a
set of WLQs previously having limited \xray\ constraints, and in this paper we
present the observational results and their implications. With these
observations, we now have complete, sensitive \xray\ coverage for the
well-defined sample of 32 WLQs with C\,{\sc iv} REW $\lesssim 15$~\AA\
defined by \citet{Ni2018}. In addition to setting tighter \xray\
constraints upon the properties of individual WLQs, these new observations
add significantly to the number of WLQs with sensitive multi-epoch \xray\
observations. They therefore also allow a search for additional examples of
extreme large-amplitude \xray\ variability events.

The layout of this paper is as follows. 
We describe the sample selection and \chandra\ observations in Section~2.
In Section~3, we detail the \xray\ photometric measurements and the derived
\xray-to-optical properties, and in Section~4 we briefly present relevant
emission-line measurements.
Section~5 presents our overall analysis results and discussion, and 
Section~6 presents a summary of our results. 
In Appendix~A, we briefly present the results from new \xmm\ 
observations of the extraordinarily luminous WLQ SDSS~J1521+5202. 

Throughout this paper, we use J2000 coordinates and a cosmology with
$H_0=67.4$~km~s$^{-1}$~Mpc$^{-1}$, $\Omega_{\rm M}=0.315$, and
$\Omega_{\Lambda}=0.685$ \citep{Planck2020}.

\section{Sample selection and \textit{Chandra} observations} \label{s-chandra}
In \citet{Ni2018}, a sample of 32 WLQs with \civ\ REW $\lesssim$ 15 \AA\ was constructed.
These WLQs were selected from radio-quiet quasars in the SDSS Data Release 7 (DR7) quasar-properties catalog of \citet{Shen2011}, including 10 of the most ``extreme'' (i.e. lowest \civ\ REW) WLQs with \civ\ REW $\lesssim$ 5 \AA.
They have $i$-band magnitude $m_i < $ 18.6 and a redshift satisfying $1.5 < z < 2.5$. 
As we want to avoid any potentially confusing X-ray absorption associated with, e.g. BAL winds, objects with BAL or mini-BAL features, narrow absorption features around \civ, or very red spectra with \gi\ > 0.45 were excluded.\footnote{\gi\ is defined as $(g-i) - (g-i)_{\rm redshift}$, where $(g-i)_{\rm redshift}$ is the median $g-i$ color of SDSS quasars at a certain redshift (see \citealt{Richards2003}).} 
Compared to other WLQ samples utilized in previous studies (e.g. \citealt{Wu2011,Wu2012, Luo2015}), this sample of WLQs has been selected in a relatively unbiased manner (no additional selection criteria such as high \civ\ blueshift and/or strong \feii/\feiii\ emission were applied) and has a relatively large sample size. Thus, it can serve as a representative sample to study the WLQ population over the full range of \civ\ REW values seen among WLQs.
WLQs in this representative sample have similar IR-to-UV SEDs (obtained from \textit{WISE}, 2MASS, SDSS, and GALEX photometry) compared with typical quasars (see \citealt{Luo2015, Ni2018} for details).

Among these 32 WLQs included in the representative sample of \citet{Ni2018}, 12 WLQs were X-ray undetected (we note that these objects were only observed with 2--5 ks \chandra\ ``snapshot'' observations at that time; e.g. see Table~2 of \citealt{Ni2018}).
Also, 2 WLQs need more secure \xray\ measurements from \chandra; previously, their X-ray properties were adopted from the \textit{XMM-Newton} slew-survey catalogue \citep{Saxton2008} and the \textit{ROSAT} all-sky survey catalogue \citep{Boller2016} that have relatively large uncertainties compared to on-axis \chandra\ detections.

\chandra\ observations for these 14 WLQs were performed between 2019 December and 2020 August, using the Advanced CCD Imaging Spectrometer \citep{Garmire2003} spectroscopic array (ACIS-S) with VF mode (which is also the mode adopted by the previous \chandra\ observations of these objects). For the 12 WLQs that were previously X-ray undetected, the exposure times were set to reach \daox\ upper limits of $\approx$ $-$0.5, if the objects remain X-ray undetected.

\section{X-ray photometric measurements and X-ray-to optical properties}
\subsection{X-ray aperture-photometry analyses} \label{ss-xray}

Following the method described in \citet{Ni2018}, we used \textit{Chandra} Interactive Analysis of Observations (CIAO) tools \citep{Fruscione2006} to process the new \chandra\ observations.
We ran the CHANDRA\_REPRO script to reprocess the observations, and then ran the DEFLARE script to remove background flares above a 3$\sigma$ level (see Table~\ref{xraytable} for the flare-cleaned exposure times for each WLQ).

From the flare-cleaned event files, we produced images for each WLQ in the 0.5--8 keV (full), 0.5--2 keV (soft), and 2--8 keV (hard) bands with the standard \textit{ASCA} grade set.
In each band, we ran WAVDETECT on the image to detect sources with a false-positive probability threshold of 10$^{-6}$.
If a source is detected in at least one band, we adopted the detected position closest to the SDSS position as the source position. 
If WAVDETECT does not detect any source within 1$''$ from the SDSS position, the SDSS position is adopted as the \hbox{X-ray} source position.

We performed aperture photometry in the soft band and the hard band for each observation. 
Source counts are extracted in the central 2$''$-radius region (which corresponds to an encircled-energy fraction of 0.959/0.907 in the soft/hard band for on-axis observations; e.g. \citealt{Luo2015}. Background counts are extracted from the annular region between the 10$''$-radius circle and the 40$''$-radius circle centered on the source position.

For each object in each band, we calculated a binomial no-source probability ($P_{B}$) from the source counts and background counts, which represents the probability of detecting the source counts by chance when no real source is present \citep[e.g.][]{Broos2007,Xue2011,Luo2015}.
For a certain band, when $P_{B}  \leqslant 0.01$, the source is considered to be detected. We thus expect $< 1$ false detection in our sample. 
The 1$\sigma$ errors of the source/background counts were obtained following \citet{Gehrels1986} that is based on Poisson and binomial statistics.
For sources considered to be undetected with $P_{B} > 0.01$, the source counts upper limits at a 90\% confidence level are derived following \citet{Kraft1991}.
We also calculated the band ratio (the ratio of hard-band counts to soft-band counts) and its uncertainty (or upper limit) utilizing the code {\sc BEHR} \citep{Park2006}.
From the band ratio, we derived the 0.5--8 keV effective power-law photon index ($\Gamma_{\rm eff}$) or its lower limit for each source with {\sc modelflux}. The intrinsic X-ray photon spectra of quasars can typically be characterized by a power-law model, $N(E) \propto E^{- \Gamma}$, and $\Gamma_{\rm eff}$ derived here can serve as a good estimate of $\Gamma$ in the absence \xray\ absorption; when \xray\ absorption is present, $\Gamma_{\rm eff}$ provides a basic estimate of spectral hardening due to the absorption.
The results are shown in Table~\ref{xraytable}.
For sources that are undetected in both the soft and hard bands, we are unable constrain $\Gamma_{\rm eff}$. 
A nominal $\Gamma_{\rm eff} = 1.4$ (which is the stacked $\Gamma_{\rm eff}$ of 30 \xray\ weak WLQs in \citealt{Luo2015}) is assumed for these sources.

We also performed the above aperture-photometry analyses when stacking all archival \chandra\ observations available and the new \chandra\ observation together for each object.
The stacked count rates are obtained with the CIAO command {\sc srcflux}.
Since WLQs may exhibit extreme \xray\ variability \citep[e.g.][]{Ni2020}, we first ensure that the count rates in all the epochs are consistent. 
As can be seen in Table~\ref{stackxraytable}, most of the WLQs do not show significant variations in count rate between the available \chandra\ epochs except for the object SDSS~J153913.47+395423.4 (hereafter SDSS~J1539+3954).
SDSS~J1539+3954 turned from an \xray\ weak WLQ to an X-ray normal WLQ between 2013 and 2019, showing an extreme X-ray flux rise by a factor of $\gtrsim 20$ \citep{Ni2020}.
According to our latest \chandra\ observation of this WLQ in 2020 June, it has now returned to being an X-ray weak WLQ (see Section~\ref{ss-xvar}).
We perform stacked aperture-photometry analyses of this object utilizing only the two X-ray weak epochs.

\subsection{X-ray to optical properties} \label{ss-aox}

We measured \aox\ values for the 12 WLQs that were previously \chandra\ undetected as well as the 2 WLQs that previously lacked high-quality \chandra\ detections.
For previously \xray\ undetected WLQs, the X-ray information utilized for calculating \aox\ is obtained from the stacked \chandra\ photometry (see Section~\ref{ss-xray}).

The X-ray-to-optical power-law slope, \aox, is defined as $0.3838 \log(L_{\rm 2~keV}/L_{2500~\mathring{\rm{A}}})$, which is equal to $0.3838 \log(f_{\rm 2~keV}/f_{2500~\mathring{\rm{A}}})$. 
In this formula, the observed flux density at rest-frame 2500~\AA, \fopt, is directly taken from the \citet{Shen2011} SDSS DR7 quasar-properties catalog.
The Galactic-absorption corrected soft-band flux (which covers \hbox{2 keV} in the rest frame for all our sources) is calculated from the net count rate in the soft band with {\sc srcflux} to derive the 2~keV flux density, $f_{\rm 2~keV}$, assuming a power-law spectrum (with $\Gamma_{\rm eff}$ measured in Section~\ref{ss-xray}) modified by Galactic absorption. 
If the source is undetected in the soft band, we calculate an upper limit for $f_{2 {\rm keV}}$ following the same method from the upper limit on the soft-band net count rate. 
The \fopt, \fx, and calculated $\alpha_{\rm OX}$ values are all listed in Table~\ref{aoxtable}.

We also derive the expected value of \aox, $\alpha_{\rm OX}(L_{2500~\mathring{\rm{A}}})$,
from the empirical \hbox{$\alpha_{\rm OX}$--$L_{\rm 2500~{\textup{\AA}}}$} relation reported in \citet{Timlin2020} for typical optically-selected quasars.\footnote{The adopted \aox-\lopt\ relation in this work is \aox\ $= (-0.199 \pm 0.011) \times {\rm log} $(\lopt) $+ (4.573\pm 0.333)$; see \citet{Timlin2020} for details.}
We then calculate $\Delta\alpha_{\rm OX}$ from the observed \aox\ and the derived $\alpha_{\rm OX}(L_{2500~\mathring{\rm{A}}})$ to measure \xray\ weakness relative to the expected \xray\ luminosity (see Table~\ref{aoxtable}).

With these updated X-ray-to-optical properties for 14 WLQs that were previously either X-ray undetected with loose X-ray upper limits or lacking high-quality \chandra\ detections, we have better constrained the X-ray-to-optical properties of the 32 WLQs in the \citet{Ni2018} WLQ representative sample.
For the 12 WLQs that were previously X-ray undetected, 7 of them of are now X-ray detected as a result of the deeper observations. We note that the X-ray flux values of these detected objects are consistent with the flux upper limits previously obtained (see Table~\ref{stackxraytable}).
For the five X-ray weak WLQs that still remain X-ray undetected (including SDSS~J1539+3954, which has transitioned into an X-ray weak state; see Section~\ref{ss-xvar} for details), we perform X-ray stacking analyses to assess their average properties, by adding
the extracted source and background counts of these objects together.
These X-ray stacking analyses (with 71.3~ks stacked exposure) show that the stacked source is still X-ray undetected, with \daox\ $< -0.74$, which corresponds to average X-ray weakness by a large factor of $> 85$.

\begin{figure}
\centering{
\includegraphics[scale=0.45]{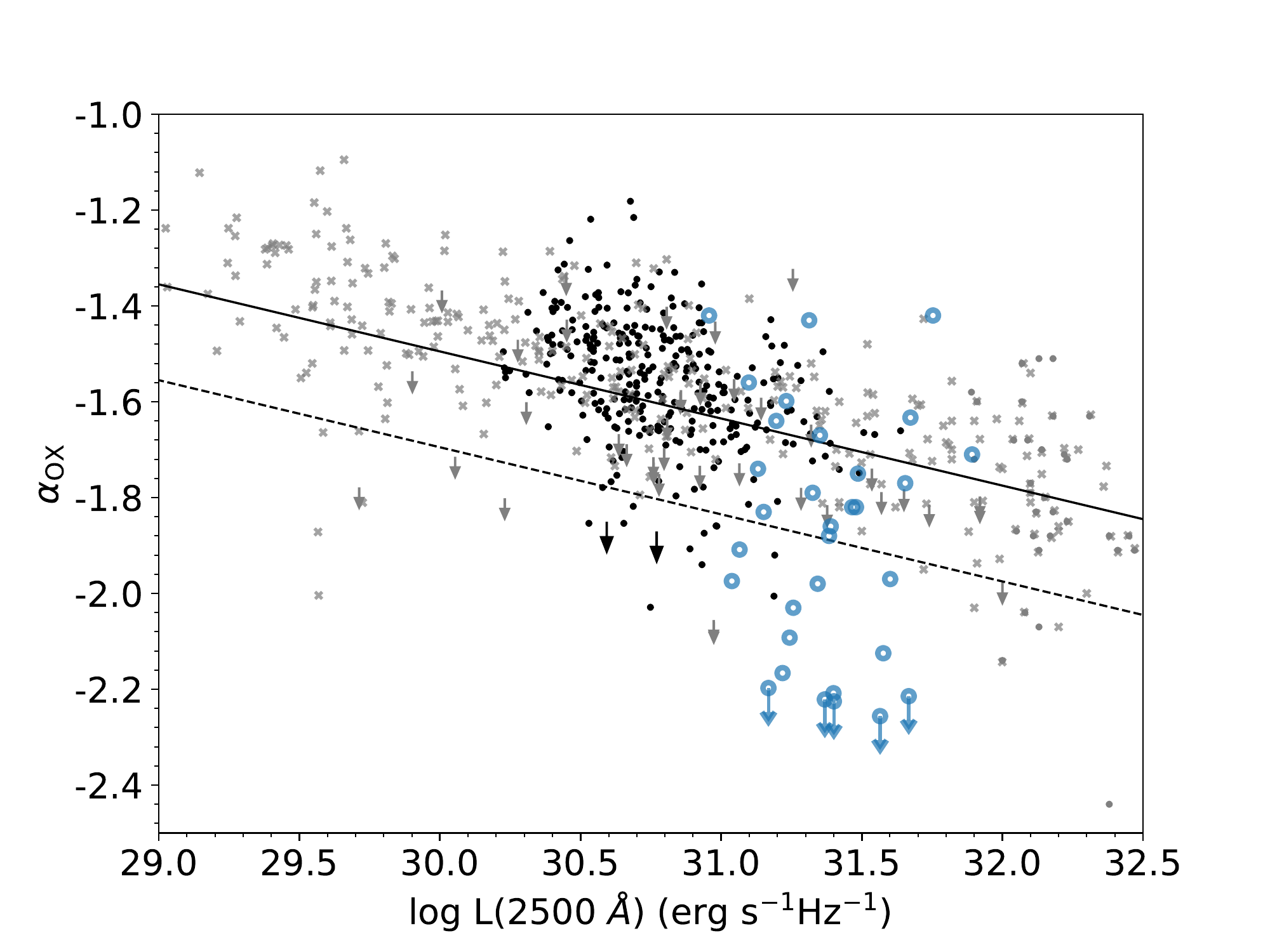}
}
\caption{X-ray-to-optical power-law slope (\aox) vs.\ \lopt\ for WLQs (blue symbols).
In cases of X-ray non-detections, the 90\% confidence upper limits of \aox\ are represented by the downward arrows.
For comparison, the \citet{Timlin2020} quasar sample that has a $\approx 99.3\%$ detection fraction (see Appendix~A of \citealt{Timlin2020}) is indicated by the black dots and downward arrows. Quasars from \citet{Steffen2006} and \citet{Just2007} are indicated by the gray crosses and downward arrows. 
The solid line shows the $\alpha_{\rm OX}$--$L_{\rm 2500~{\textup{\AA}}}$ relation from \citet{Timlin2020}; 
the dashed line ($\Delta\alpha_{\rm OX}=-0.2$) represents the adopted division between \xray\ normal and \xray\ weak quasars in this study, which corresponds to a $\approx 1.3\sigma$
($\approx 90\%$ single-sided confidence level) offset given the \daox\ distribution (e.g. \citealt{Steffen2006, Luo2015}).
}
\label{alphaox}
\end{figure}

\begin{figure}
\centering{
\includegraphics[scale=0.45]{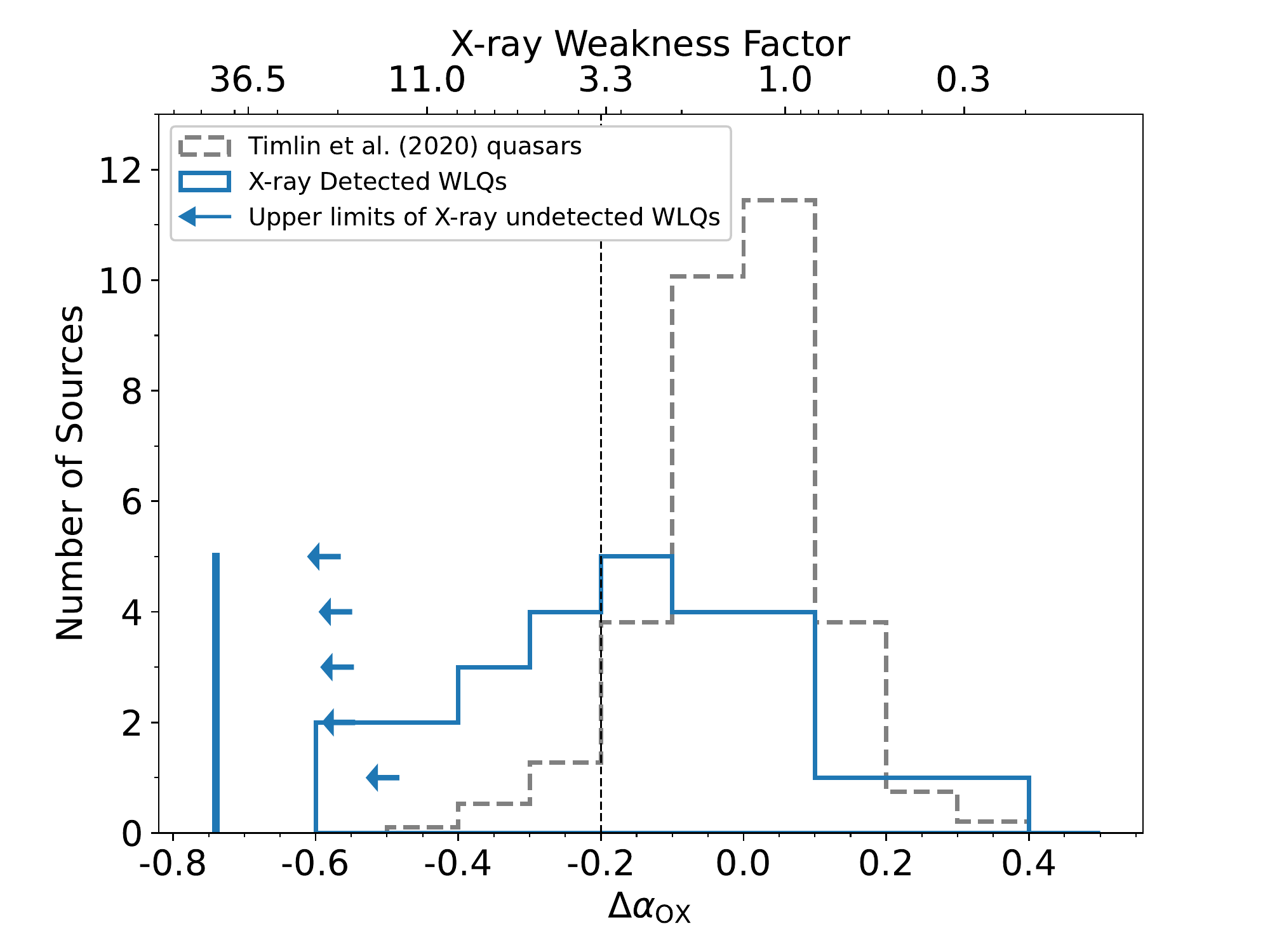}
}
\caption{Distribution of $\Delta\alpha_{\rm OX}$ values. 
The solid histogram represents the \daox\ distribution of the 27 WLQs in the representative sample that are X-ray detected.
The leftward arrows show the 90\% confidence $\Delta\alpha_{\rm OX}$ upper limits for five X-ray non-detected WLQs.
The vertical bar represents the stacked \daox\ upper limit for these X-ray non-detected WLQs.
The dashed histogram represents the $\Delta\alpha_{\rm OX}$ distribution for the \citet{Timlin2020} quasar sample that has a $\approx 99.3\%$ detection fraction, which is plotted for comparison (scaled to have the same apparent number of sources as the WLQ sample).
The vertical dashed line (at $\Delta\alpha_{\rm OX}=-0.2$) represents the adopted threshold for defining \xray\ weakness in this study.
}
\label{dalphaox}
\end{figure}

The distributions of $\alpha_{\rm OX}$ and $\Delta\alpha_{\rm OX}$ for this representative sample that consists of 32 objects are shown in Figures~\ref{alphaox} and \ref{dalphaox}.
We also compare the \daox\ distribution of WLQs with the \daox\ distribution of typical quasars from Appendix A of \citet{Timlin2020} with the Anderson-Darling test.\footnote{We note that the Peto-Prentice test in the Astronomical Survival Analysis package (ASURV; e.g. \citealt{Feigelson1985,Lavalley1992}) that is often adopted for censored data assumes that the censored data follow the same intrinsic distribution as the uncensored data, which is not a realistic assumption in our case (see Figures~\ref{alphaox} and \ref{dalphaox}). Therefore, we adopted a Monte Carlo approach as described in the main text, and used the Anderson-Darling test for statistical evaluation in this work.}
This \citet{Timlin2020} ``high X-ray detection fraction'' quasar sample consists of 304 SDSS quasars (which are unobscured, broad-line AGNs) at $z = 1.7$--2.7 with $i$-band magnitude $<$ 20.2 that have serendipitous sensitive \chandra\ X-ray observations (the detection fraction is $\approx 99.3\%$), with BAL and red quasars removed.

Since the Anderson-Darling test cannot properly utilize censored data, for 2 out of 304 quasars in the \citet{Timlin2020} high-detection-fraction sample that only have \daox\ upper limits, their \daox\ values utilized in the analyses are drawn from the probability density function of the \daox\ values of other \xray\ detected objects in the sample (the maximum value allowed to be drawn is set at the upper-limit value). For the 5 WLQs that are \xray\ undetected, their \daox\ values are drawn from a normal distribution centered at their stacked \daox\ limit with a scatter of 0.1, with the maximum value allowed to be drawn as the upper-limit value.
This procedure is repeated 1000 times to obtain an average test result, and this Monte Carlo approach is adopted throughout this work to account for the \daox\ upper limits of typical quasars/WLQs when performing statistical tests (see Section~\ref{ss-daoxciv}).
We found that the \daox\ distributions of WLQs and typical quasars are different at a $\approx 4.5 \sigma$ level. 
As we have already taken luminosity effects into account by converting \aox\ to \daox, our comparison should be valid between samples with different luminosity ranges. We also note that this test result does not change materially when we limit our analyses to the bright objects in the \citet{Timlin2020} high-detection-fraction sample with luminosities similar to those of WLQs in the representative sample.
Among X-ray normal (or \xray\ strong; $\Delta\alpha_{\rm OX} \geqslant -0.2$) objects, the \daox\ distributions of WLQs and typical quasars are not significantly different ($P_{\rm null} = 0.26$).
Among X-ray weak ($\Delta\alpha_{\rm OX }< -0.2$) objects, the \daox\ distributions of WLQs and typical quasars are different at a $\approx 2.9 \sigma$ level.
This indicates that WLQs differ from typical quasars mainly in having a strong tail toward X-ray weakness, as expected from the TDO model for WLQs.

We also use X-ray stacking analyses to derive the average spectral properties of X-ray weak WLQs and X-ray normal WLQs in the representative sample. The stacked $\Gamma_{\rm eff}$ is $1.1^{+0.2}_{-0.1}$ for X-ray weak WLQs, and $1.8^{+0.1}_{-0.1}$ for X-ray normal WLQs. 
These results further support the high apparent levels of intrinsic X-ray absorption, Compton reflection, and/or scattering among X-ray weak WLQs, supporting the existence of shielding materials with large column densities along our line sight, which we proposed to be the TDO (e.g. \citealt{Luo2015, Ni2018}).

\section{Emission-line measurements}

The \civ\ REWs of WLQs in the representative sample have been measured in Section~4.3 of \citet{Ni2018}; \civ\ REW has been widely adopted to indicate the number of ionizing EUV photons reaching the high-ionization BLR.
Since the strength of \heii\ $\lambda$1640 emission has also been shown to be an effective tracer of the number of EUV photons available (and it is a ``cleaner'' tracer compared to the \civ\ emission strength; e.g.\ \citealt{Timlin2021} and references therein), we also measure the REWs of \heii\ emission for WLQs in the representative sample following the method of \citet{Timlin2021}.
The local continuum in the \heii\ emission region was obtained by fitting a linear model to the median values in the continuum windows 1420--1460 \AA\
and 1680--1700 \AA. 
The \heii\ REW was then measured by integrating directly the continuum-normalized flux in the window 1620--1650 \AA.
The errors of \heii\ REW are measured by perturbing the fluxes using the flux errors (see \citealt{Timlin2021} for details).
The median \heii\ REW error of WLQs in our sample is $\approx 0.6$ \AA.
If the \heii\ emission line is not detected significantly, an upper limit was estimated.
The results can be seen in Table~\ref{heiitable}.
We note that most of the WLQs do not have \heii\ emission detected; only upper limits for \heii\ REW are obtained.
We also note that, as WLQs have weak and sometimes blueshifted high-ionization emission lines, there might be uncertainties associated with their redshift measurements. We verified that our \heii\ measurement results do not change materially when different redshift measurements are adopted (e.g. \citealt{Hewett2010, Schneider2010}).

We also measure the \heii\ REW limits in the stacked spectrum of X-ray weak/normal WLQs in our sample.
Each WLQ spectrum is normalized to its median flux value in the range 1700--1705 \AA\ before being stacked together (see \citealt{Timlin2021} for details about the stacking process). For the 16 \xray\ weak WLQs in the representative sample, the stacked \heii\ REW upper limit is 0.12~\AA. For the 16 \xray\ normal WLQs, the stacked \heii\ REW upper limit is 0.29~\AA.

\section{Analysis results and discussion}

\subsection{Assessing and refining the TDO model for WLQs} \label{ss-tdo}
The X-ray-to-optical properties of the representative WLQ sample as presented in Section~\ref{ss-aox} can give us some basic constraints on the nature of the TDO we proposed to explain the X-ray and multiwavelength properties of WLQs.
From the \daox\ distribution of WLQs in our sample, we can infer that the TDO, if present, has an average global covering factor of $\approx 0.5$ among WLQs.
The TDO should also be able to produce X-ray weakness by an average factor of $> 85$ for $\approx 15\%$ of the WLQs, and by an average factor of $\approx$ 9 for $\approx 35\%$ of the WLQs, suggesting that the shielding column density could range from $N_{\rm H}$ $\sim 10^{23-24}~{\rm cm}^{-2}$ to $N_{\rm H}$ $\gtrsim 10^{24}~{\rm cm}^{-2}$. 
The X-ray weak WLQs in our sample have a stacked $\Gamma_{\rm eff} \approx 1.1$ and a stacked X-ray weakness factor of $\approx$~13. 
To produce this level of X-ray weakness with a simple intrinsic absorption model, a TDO with $N_{\rm H}$ $\sim 5 \times 10^{23}~{\rm cm}^{-2}$ is required. 
However, such a TDO will produce an \xray\ spectrum with $\Gamma_{\rm eff} \sim 0.3$, which is not consistent with the stacked $\Gamma_{\rm eff}$ $\approx 1.1$ we obtained, suggesting that more complex absorption (and/or Compton reflection/scattering or a soft-excess component) is present, and that how the TDO modifies the X-ray properties varies from object-to-object.
The complex absorption (and/or Compton-reflection/scattering) effects caused by the TDO are also indicated by the lack of bimodality in the \daox\ distribution of WLQs. If the TDO had a rather simple nature so that similar \xray\ ``blocking'' effects were present among all X-ray weak WLQs, we should be able to observe a ``clustering'' of \daox\ on the X-ray weak side, rather than the long tail toward X-ray weakness observed in Figure~\ref{dalphaox}.

We note that there are also other possible explanations for the distinctive X-ray properties of WLQs listed in Section~1, such as intrinsic differences/variations in the coronal emission, or gravitational light-bending that leads to different amounts of X-ray emission reaching the observer when the distances between the corona and the central black hole are different (e.g. \citealt{Miniutti2004}).
However, neither of these scenarios can explain why the distinctive \xray\ properties are only observed among WLQs rather than quasars with typical optical/UV emission-line properties -- the link between these scenarios and weak emission lines is not clear. These scenarios also cannot explain the high apparent level of X-ray absorption we observed among X-ray weak WLQs. 

Furthermore, while shielding materials located on a much larger (by a factor of $\gtrsim$ 100) scale might explain the X-ray properties of WLQs (e.g. ``clumps'' near the torus region), the cause for the defining weak high-ionization emission lines would remain unexplained. WLQs have IR-to-UV SEDs similar to those of typical quasars \citep[e.g.][]{Luo2015, Ni2018}, so the weak high-ionization emission lines cannot be attributed to any obvious difference in the continuum level.
As only shielding materials located \textit{between} the central X-ray source and the high-ionization BLR can naturally explain all the multiwavelength properties of WLQs, the TDO model (which includes both a thick disk and an outflow component) seems to be the most viable available solution.

\subsection{\daox-\civ\ REW and \daox-\heii\ REW distributions of WLQs vs. typical quasars} \label{ss-daoxciv}

The relations between \daox\ and \civ\ REW or \heii\ REW among typical quasars reflect the link between the X-ray emission strength and the strength of the ionizing EUV continuum (e.g. \citealt{Timlin2021} and references therein). 
While EUV photons may be generated from a ``warm corona" in the inner-disk region via Comptonization (e.g. \citealt{Petrucci2018}) and X-ray photons are thought to be produced by the ``hot corona'' in the vicinity of the central black hole, these relations indicate that there is a strong coupling between the X-ray emission and EUV emission.
The unusual X-ray properties of WLQs and their weak emission lines naturally lead one to wonder whether the relation between \daox\ and \civ\ or \heii\ REW also applies to WLQs, and whether this may challenge the universal existence of the coupling between the X-ray emission and the EUV emission among quasars.

\begin{figure*}
\centering{
\includegraphics[scale=0.4]{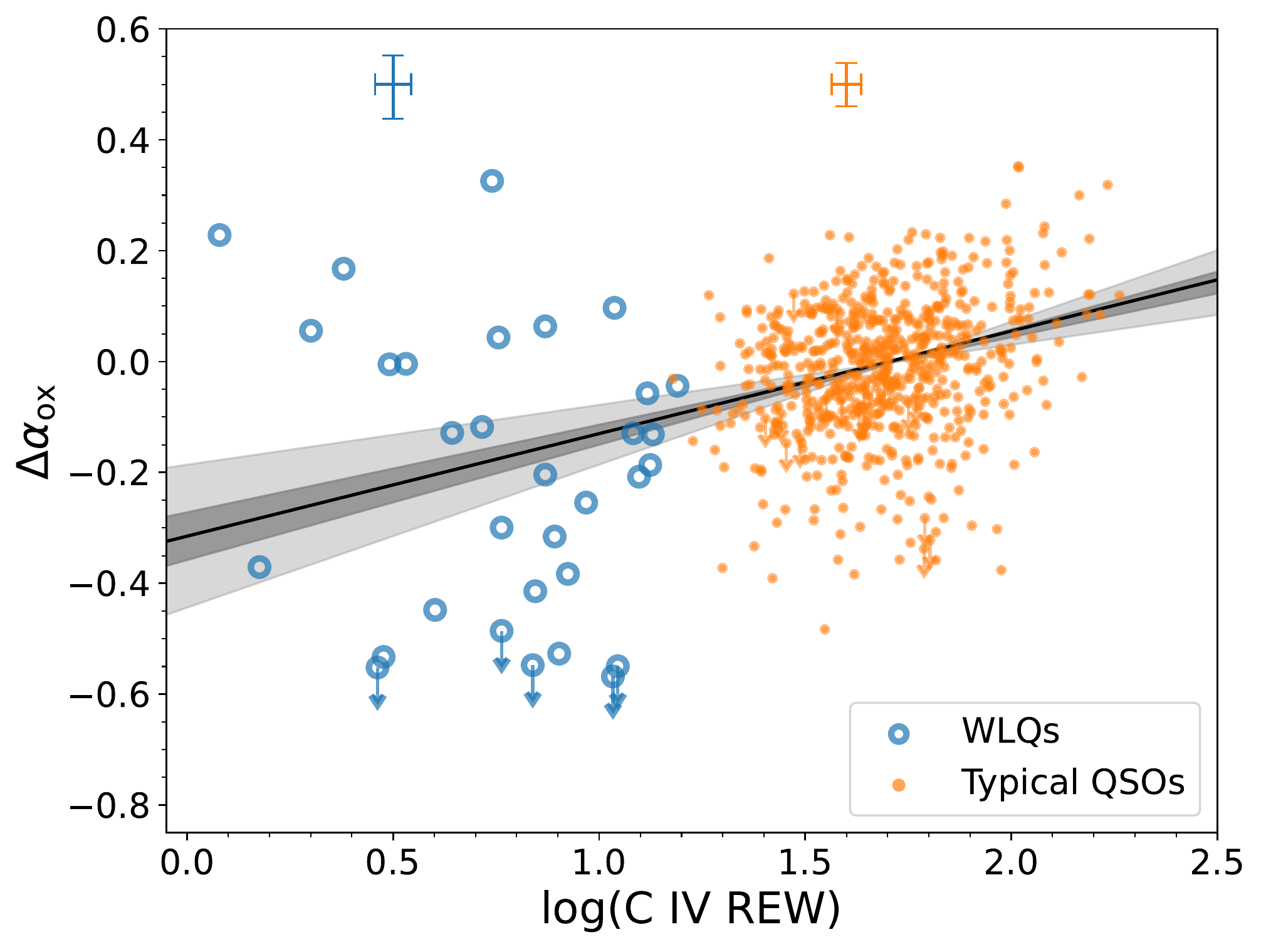}
\includegraphics[scale=0.4]{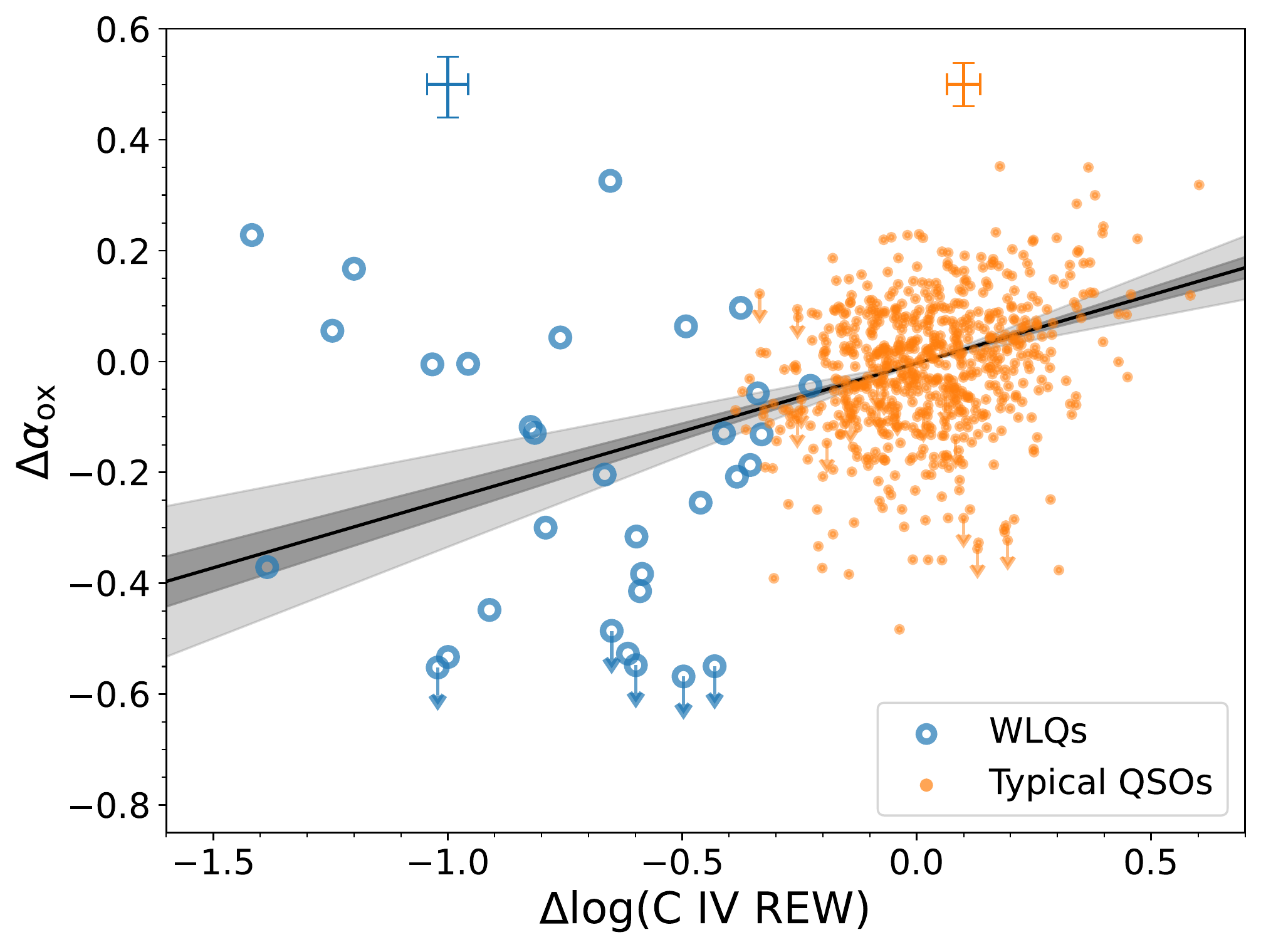}}
\caption{\textit{Left:} \daox\ vs. log~\civ~REW (in units of \AA) for WLQs in the representative sample as well as typical quasars in \citet{Timlin2020}. 
\textit{Right:} \daox\ vs. $\Delta$log~\civ~REW for WLQs in the representative sample as well as typical quasars in \citet{Timlin2020}. 
The best-fit \daox-\civ\ REW (\daox-$\Delta$log~\civ\ REW) relation \citep{Timlin2020} is shown as the black solid line, with its 1$\sigma$/3$\sigma$ confidence intervals shown as the dark/light gray-shaded region in the left (right) panel.
The median measurement errors of \daox\ and \civ\ REW for WLQs and typical quasars are shown as error bars at the top of each panel.
}
\label{daoxciv}
\end{figure*}

In the \daox\ vs. log \civ~REW space, WLQs appear to show relatively larger scatter compared to typical quasars in \citet{Timlin2020} (see the left panel of Figure~\ref{daoxciv}).
To test whether the deviations of WLQs from the \daox-\civ~REW relation are statistically different compared to those of typical quasars, we use the Anderson-Darling test with the Monte Carlo procedure as described in Section~\ref{ss-aox}.
All 32 WLQs and 637 quasars in the \citet{Timlin2020} sample that have \civ\ detections are utilized in the Anderson-Darling test, including 5 WLQs and 18 quasars in the \citet{Timlin2020} sample that only have \daox\ upper limits.\footnote{This \civ-detected subsample of the \citet{Timlin2020} serendipitous quasar sample selects quasars with spectral signal-to-noise ratio  $\geqslant 3$, and quasars in this subsample have properties similar to the whole serendipitous quasar sample (see Table~1 of \citealt{Timlin2020}).}

We found that the difference in the distributions of \daox\ residuals from the best-fit \daox-log~\civ~REW relation reported in \cite{Timlin2020} (as can be seen in the left panel of Figure~\ref{daoxciv}) among WLQs and typical quasars is significant at a $\approx 4.0~\sigma$ level.

To test whether the observed phenomenon is influenced by the Baldwin effect (which shows that \civ\ REW is strongly linked with \lopt; e.g. \citealt{Baldwin1977}), we perform the above analyses for \daox\ and $\Delta$log~\civ~REW (which is calculated as log \civ\ REW minus the expected log \civ\ REW value from the \civ\ REW-\lopt\ relation reported in \citealt{Timlin2020}).
The best-fit \daox-$\Delta$log~\civ\ REW relation is derived utilizing the Python package {\sc linmix} \citep{Kelly2007}.
We found that the difference in the distributions of \daox\ residuals from the \daox-$\Delta$log~\civ\ REW relation (as can be seen in the right panel of Figure~\ref{daoxciv}) between WLQs and typical quasars is still significant at a $\approx 4.0~\sigma$ level according to the Anderson-Darling test.
The above analysis results do not change qualitatively when we limit our comparisons to WLQs and the brightest 64 objects among the \cite{Timlin2020} quasars with \civ\ detections, which have similar luminosities to the WLQs in our representative sample and are all X-ray detected. The above analysis results also do not change qualitatively when we perturb the \daox\ values with the measurement errors.

\begin{figure*}
\centering{
\includegraphics[scale=0.4]{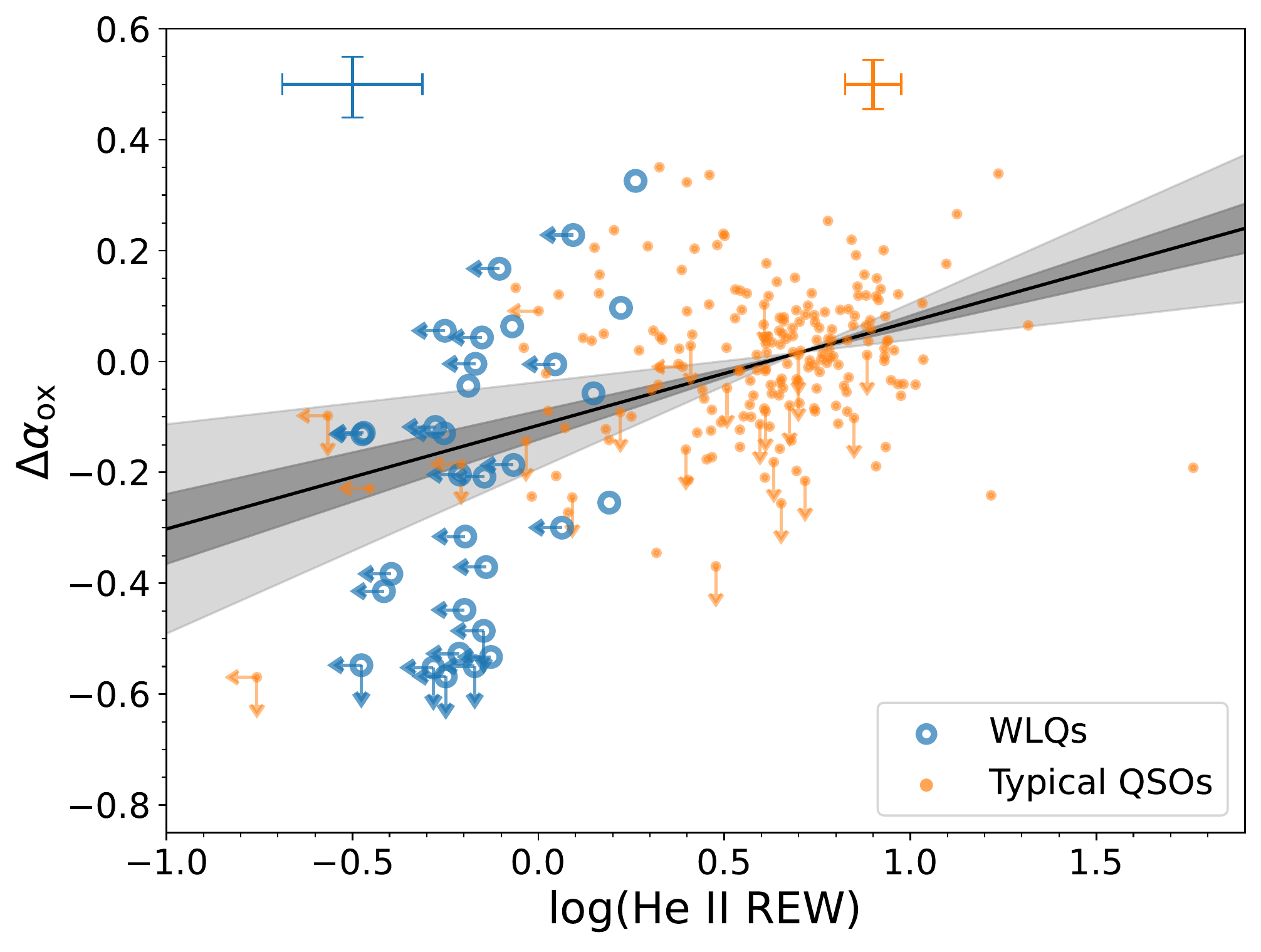}
\includegraphics[scale=0.42]{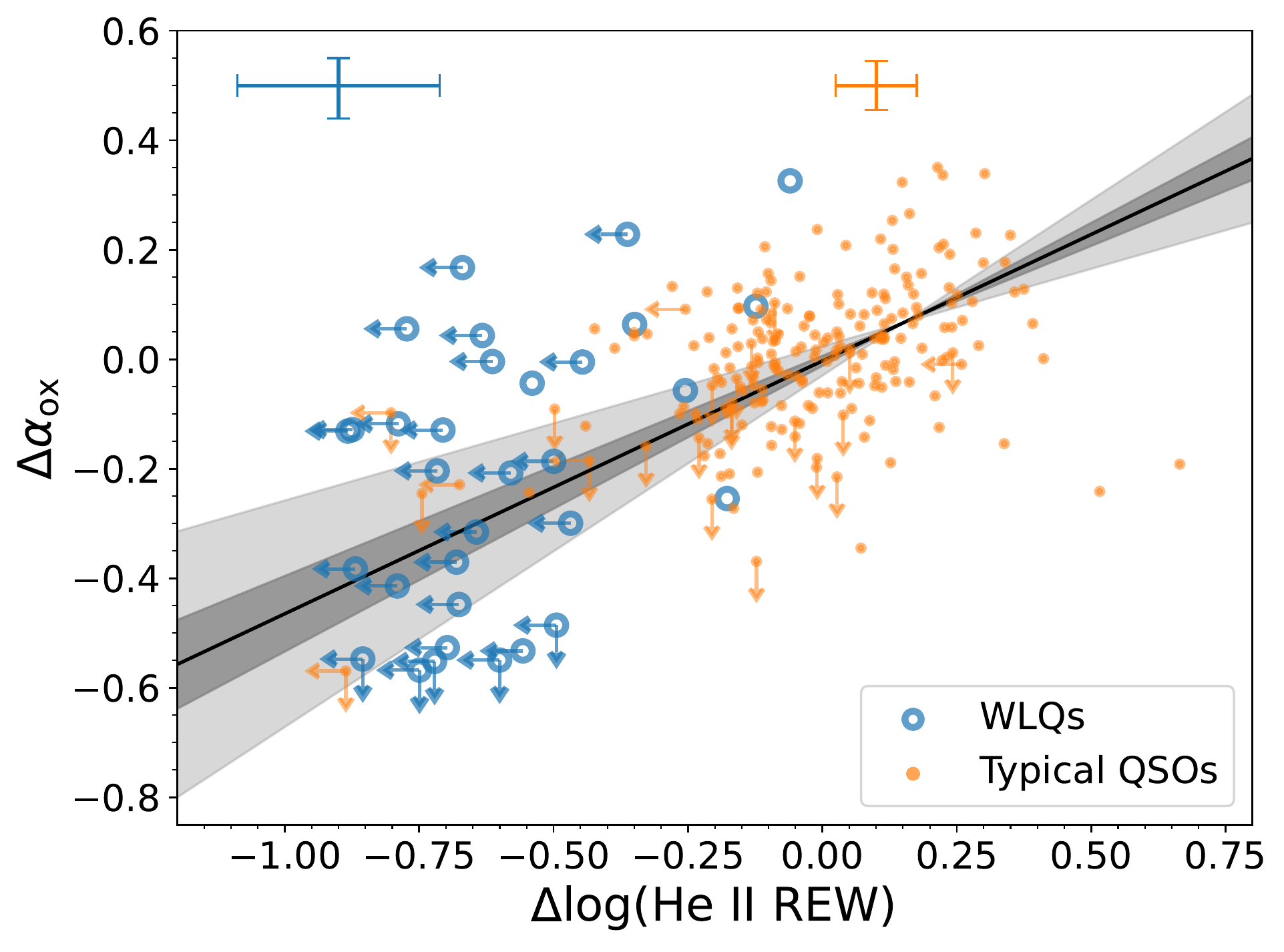}}
\caption{ \textit{Left:} \daox\ vs. log~\heii~REW (in units of \AA) for WLQs in the representative sample as well as typical quasars in \citet{Timlin2021}. 
\textit{Right:} \daox\ vs. $\Delta$log~\heii~REW for WLQs in the representative sample as well as typical quasars in \citet{Timlin2021}. 
The best-fit \daox-\heii\ REW (\daox-$\Delta$log~\heii\ REW) relation reported in \citet{Timlin2021} is shown as the black solid line, with its 1$\sigma$/3$\sigma$ confidence intervals shown as the dark/light gray-shaded region in the left (right) panel.
The median measurement errors of \daox\ and \heii\ REW for WLQs and typical quasars are shown as error bars at the top of each panel.
}
\label{daoxhe}
\end{figure*}

We also use the Peto-Prentice test with a Monte Carlo approach to assess if the deviations of WLQs and typical quasars from the \daox-\heii\ REW relation derived in \citet{Timlin2021} are different (see the left panel of Figure~\ref{daoxhe}); the Peto-Prentice test can incorporate the censored measurements of \heii\ into the analyses.
For 5 WLQs and 21 out of 206 quasars in the \citet{Timlin2021} sample that only have \daox\ upper limits, their \daox\ values utilized in the analysis are drawn following the method described in Section~\ref{ss-aox}.
We can see that in the \daox\ vs. log \heii~REW space, WLQs exhibit obviously larger scatter compared to typical quasars.
We found that the difference in the distributions of residuals from the \daox-\heii~REW relation between WLQs and typical quasars is significant at $\approx 5.6\sigma$ according to the Peto-Prentice test.

It has also been found that for typical quasars, \daox\ follows a tight correlation with the difference between log values of measured \heii\ REWs and expected \heii\ REWs from the \heii\ REW-\lopt\ relation ($\Delta$log~\heii~REW; \citealt{Timlin2021}), as shown by the black line in the right panel of Figure~\ref{daoxhe}.
However, WLQs in our representative sample do not follow this relation in a similar pattern as that of typical quasars, again showing larger scatter.
We also use the Peto-Prentice test to assess statistically the difference in the residuals of \daox\ values compared to the expectation from the \daox-$\Delta$log~\heii~REW relation between WLQs and typical quasars.
We found that this difference is significant at $\approx 7.8\sigma$.
The above analysis results do not change qualitatively when we limit our comparisons to WLQs and the brightest 64 objects among the \cite{Timlin2021} quasars, which have similar luminosities to the WLQs in our representative sample with an X-ray detection fraction of $\approx 92\%$. The above analysis results also do not change qualitatively when we perturb the \daox\ values with the measurement errors.

The statistical test results above clearly indicate that, in the \daox\ vs. high-ionization emission-line strength space, WLQs do not behave in a manner similar to that of typical quasars (WLQs have larger scatter). 
However, this does not necessarily mean that the link between \xray\ emission and EUV emission that exists among the general quasar population is not applicable to WLQs, as \civ\ REW or \heii\ REW can only serve as an indicator of the number of EUV photons that reach the high-ionization BLR (where the relevant emission lines are produced) rather than the total number of EUV photons emitted, and \daox\ can only measure the relative strength of the X-ray emission that has successfully reached the observer.

The TDO model is plausibly able to explain why the distribution of WLQs' deviations from the \daox-\civ\ (or \heii) REW relation is broader than that of typical quasars, while the coupling between intrinsic X-ray emission and EUV emission still holds.
In the context of the TDO model, the ionizing EUV photons produced are largely prevented from reaching the BLR due to shielding by the TDO (e.g. see Figure~1 of \citealt{Ni2018}).
For X-ray normal WLQs, if their EUV emission were not heavily shielded by the TDO, their \civ/\heii\ REWs should not be as small as observed.
According to the correlation between \daox\ and \civ\ (or \heii) REW, we expect them to have \daox\ values similar to those of typical quasars, which is consistent with their observed X-ray emission strength (see Section~\ref{ss-aox}).
For \xray\ weak WLQs, not only is their EUV emission strength not represented by their \civ/\heii\ REWs, but also their \daox\ values cannot serve as measurements of the intrinsic X-ray emission strength. 
The \daox\ values of X-ray weak WLQs reflect the observed X-ray emission strength with the presence of the TDO along the line of sight, which blocks the intrinsic X-ray emission with heavy absorption. 
Thus, these X-ray weak WLQs fall below the \daox-\civ\ REW (or \heii\ REW) relation when we ``relocate'' them to have \civ/\heii\ REWs similar to those of typical quasars.
When we consider the WLQ population altogether, the significantly larger scatter of WLQs compared to that of typical quasars in Figures~\ref{daoxciv} and \ref{daoxhe} is expected, as the X-ray weak WLQs are ``moved'' downward along the \daox\ axis by the TDO.
The larger scatters of WLQs compared to typical quasars in the \daox\ vs. \civ\ (or \heii) REW space lead to the significant test results when comparing the deviations of these two samples from the best-fit relation.
As the \heii\ REW serves as a better tracer of EUV ionizing photons (e.g. see \citealt{Timlin2021} and references therein), typical quasars have a smaller scatter in the \daox\ vs. \heii\ REW space compared to the \daox\ vs. \civ\ REW space. Thus, the test results are more significant in the \daox\ vs. \heii\ REW space.

\subsection{Spectral tracers of \xray\ weakness among WLQs} \label{ss-tracer}

\begin{figure}
\centering{
\includegraphics[scale=0.45]{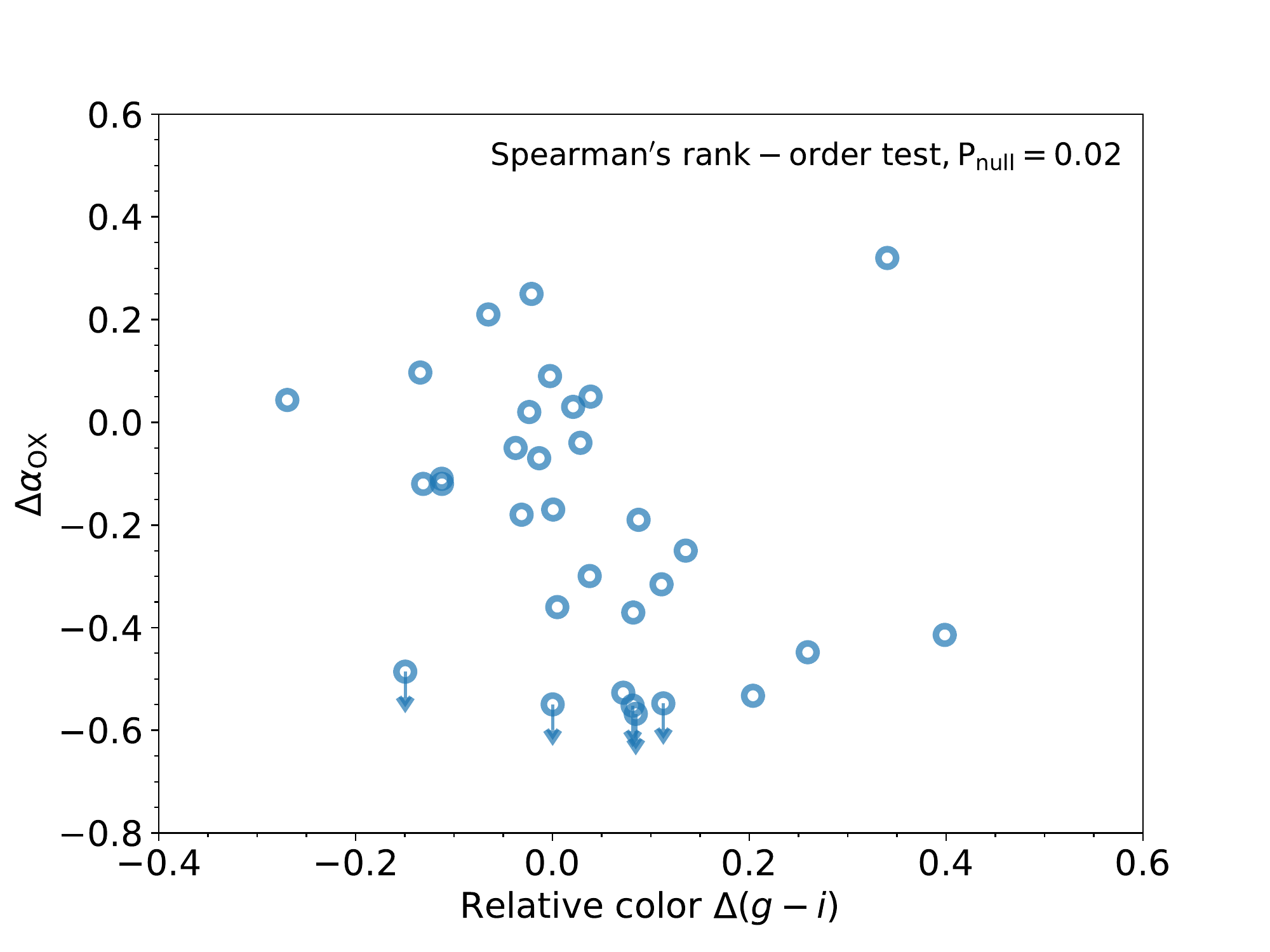}}
\caption{\daox\ vs. relative color for WLQs in the representative sample.
For X-ray non-detections, the 90\% confidence upper limits of $\Delta\alpha_{\rm OX}$ are marked as the downward arrows.
}
\label{daoxgi}
\end{figure}

\begin{figure}
\centering{
\includegraphics[scale=0.45]{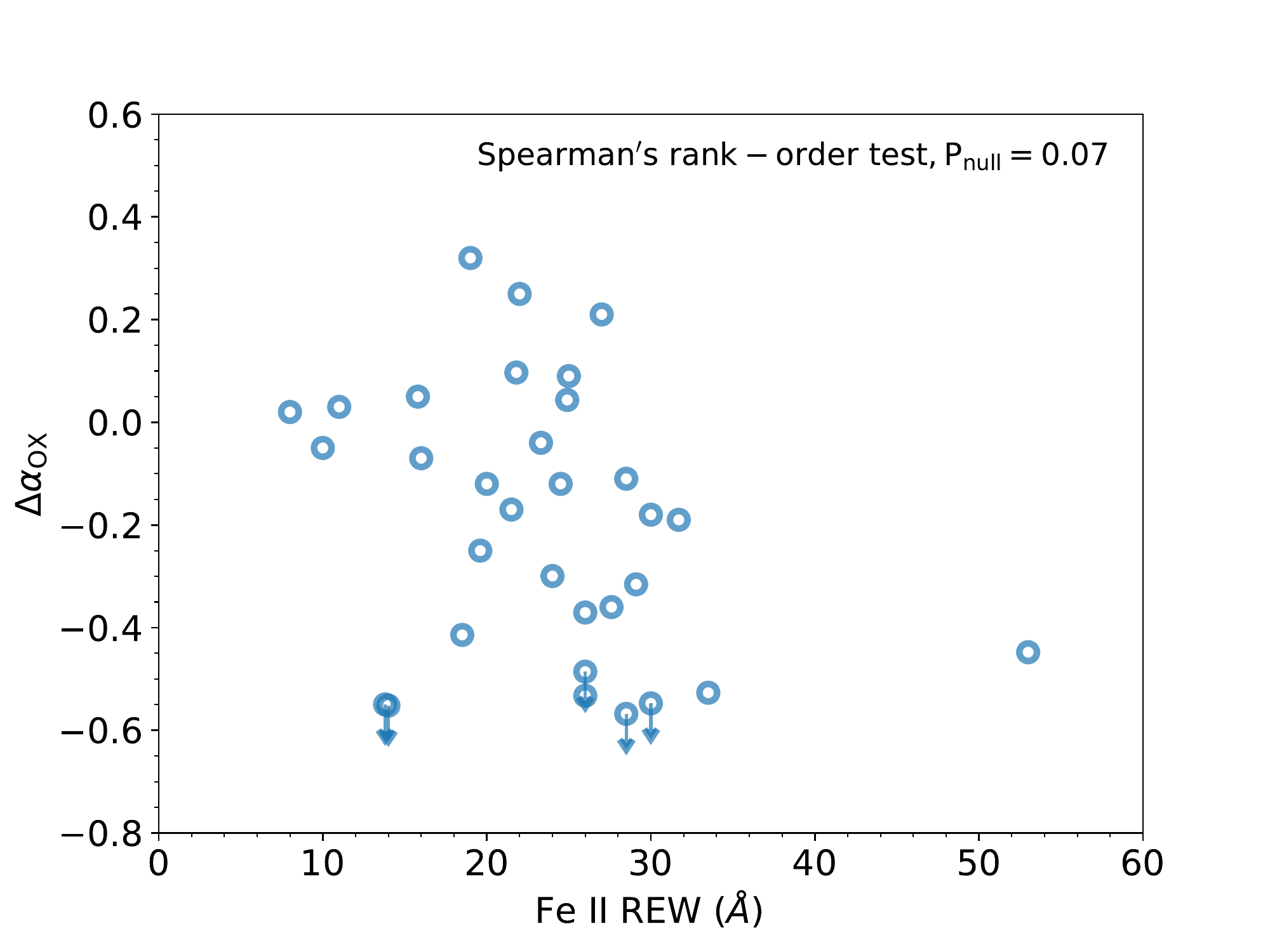}}
\caption{\daox\ vs. \feii\ REW for WLQs in the representative sample.
For X-ray non-detections, the 90\% confidence upper limits of $\Delta\alpha_{\rm OX}$ are marked as the downward arrows.
}
\label{daoxfeii}
\end{figure}

In \citet{Ni2018}, we identified \gi\ and \feii\ REW as potential tracers of X-ray weakness among WLQs.
With 7 additional X-ray detections and the improved \daox\ constraints for objects that remain X-ray undetected, these conclusions still hold, with test statistics roughly unchanged compared to \citet{Ni2018}. The significance levels of the correlations between \daox\ and \gi\ or \feii\ REW are close to, but below, 3$\sigma$ for the WLQ representative sample.
The details are shown in Figures~\ref{daoxgi} and \ref{daoxfeii}.
Both relations have considerable scatter.
We also note that the combination of \gi\ and \feii\ is not able to give a significantly better prediction of \daox\ than either of these quantities individually.

We note that for the Full sample in \citet{Ni2018} that has 63 WLQs, both \gi\ and \feii\ REW correlate with \daox\ significantly.
However, around half of the objects in the Full sample were not selected in an unbiased manner; they were selected for \chandra\ observations with additional requirements such as large \civ\ blueshifts and strong \feii/\feiii\ emission. Thus, while both \gi\ and \feii\ REW have the potential to predict X-ray weakness, we still need a larger sample of WLQs that have been selected in an unbiased manner to confirm it. 
This will help to clarify the nature of the shielding material among WLQs, which we propose to be the TDO.
In the context of the TDO model, whether a WLQ is observed as X-ray weak or X-ray normal is largely determined by its inclination angle. 
If the relation between \daox\ and \gi\ is confirmed, it might be explained if the amount of dust tends to increase toward the equatorial plane (e.g. see \citealt{Elvis2012, Luo2015} for details). This would cause mild excess reddening when quasars are viewed at large inclination angles, so that X-ray weak WLQs have redder colors compared to those of X-ray normal WLQs. If the relation between \daox\ and \feii\ REW is confirmed, it may be a consequence of aspect-dependent effects of the disk emission (e.g. \citealt{Wang2014}).

\subsection{X-ray variability of WLQs} \label{ss-xvar}
\begin{figure}
\centering{
\includegraphics[scale=0.5]{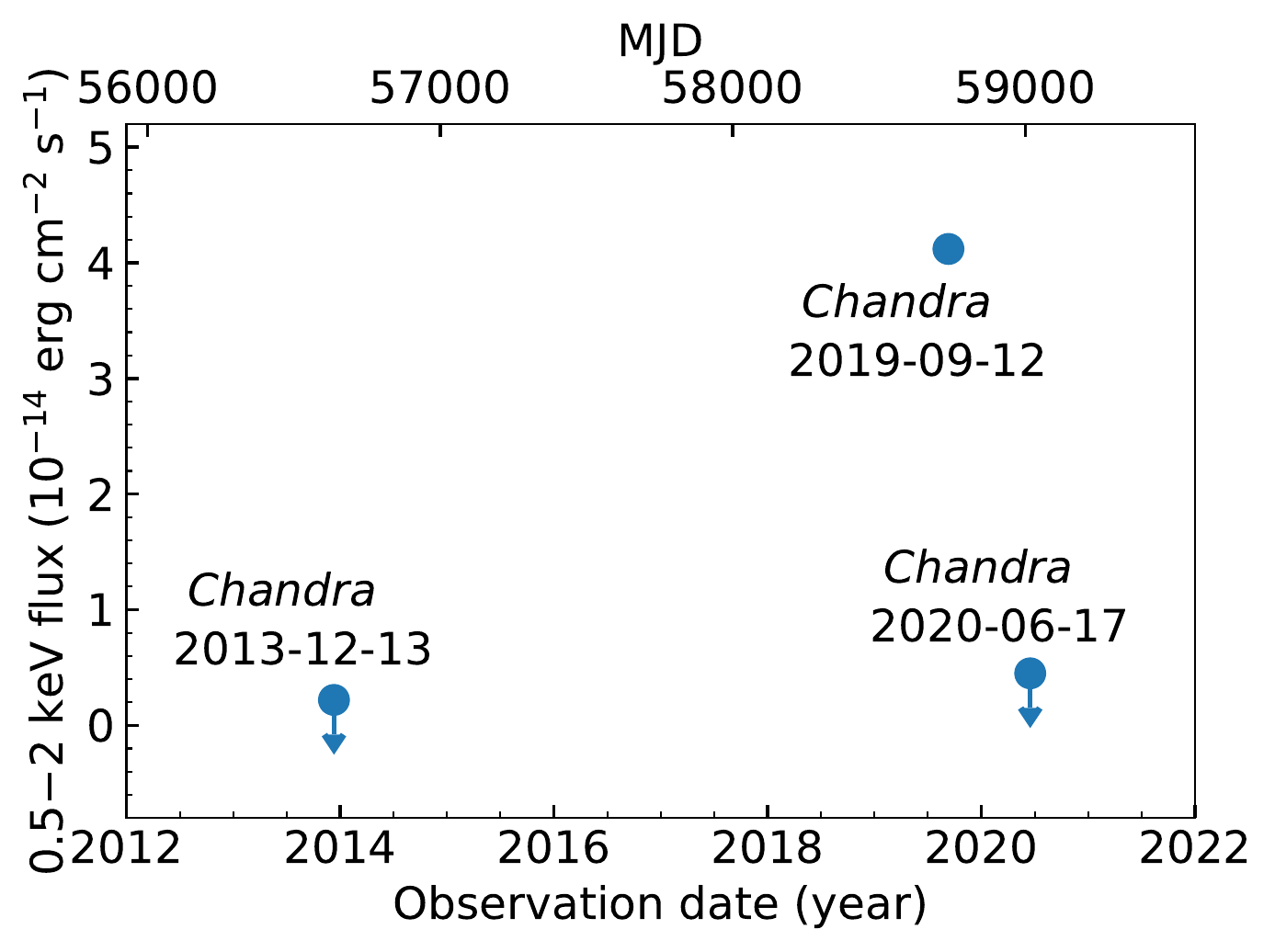}}
\caption{The historical 0.5--2 keV X-ray flux of SDSS~J1539+3954. The solid circle represents the detected flux; downward arrows represent the upper limits on flux obtained from a non-detection.
This quasar rose in \xray\ flux by a factor of $\gtrsim 20$ \citep{Ni2020} and then declined by a factor of $\gtrsim 9$.
}
\label{xlc}
\end{figure}

\begin{figure*}
\centering
\includegraphics[scale=0.45]{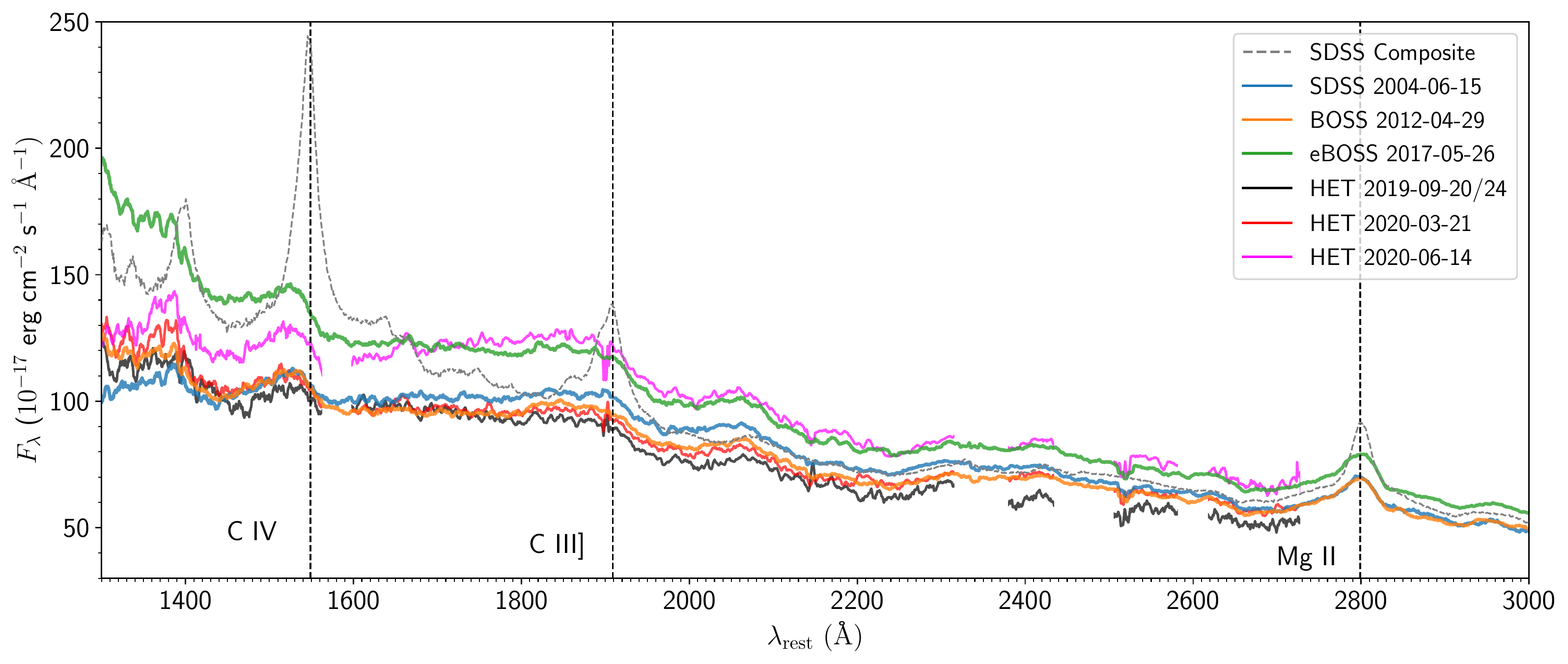}
\caption{Spectra of SDSS~J1539+3954 taken at different epochs. 
For HET spectra, we mask the areas suffering from channel discontinuities and telluric absorption (G. Zeimann 2019, private communication). 
The SDSS quasar composite spectrum from \citet{VB2001} is scaled to the 2004 SDSS spectrum of SDSS~J1539+3954 at rest-frame 2240 \AA\ and plotted in the background for comparison.
Note that the emission lines (e.g. \civ) remained weak despite the strong X-ray flux variations occurring contemporaneously.
}
\label{hetspec}
\end{figure*}

\begin{figure*}
\centering
\includegraphics[scale=0.45]{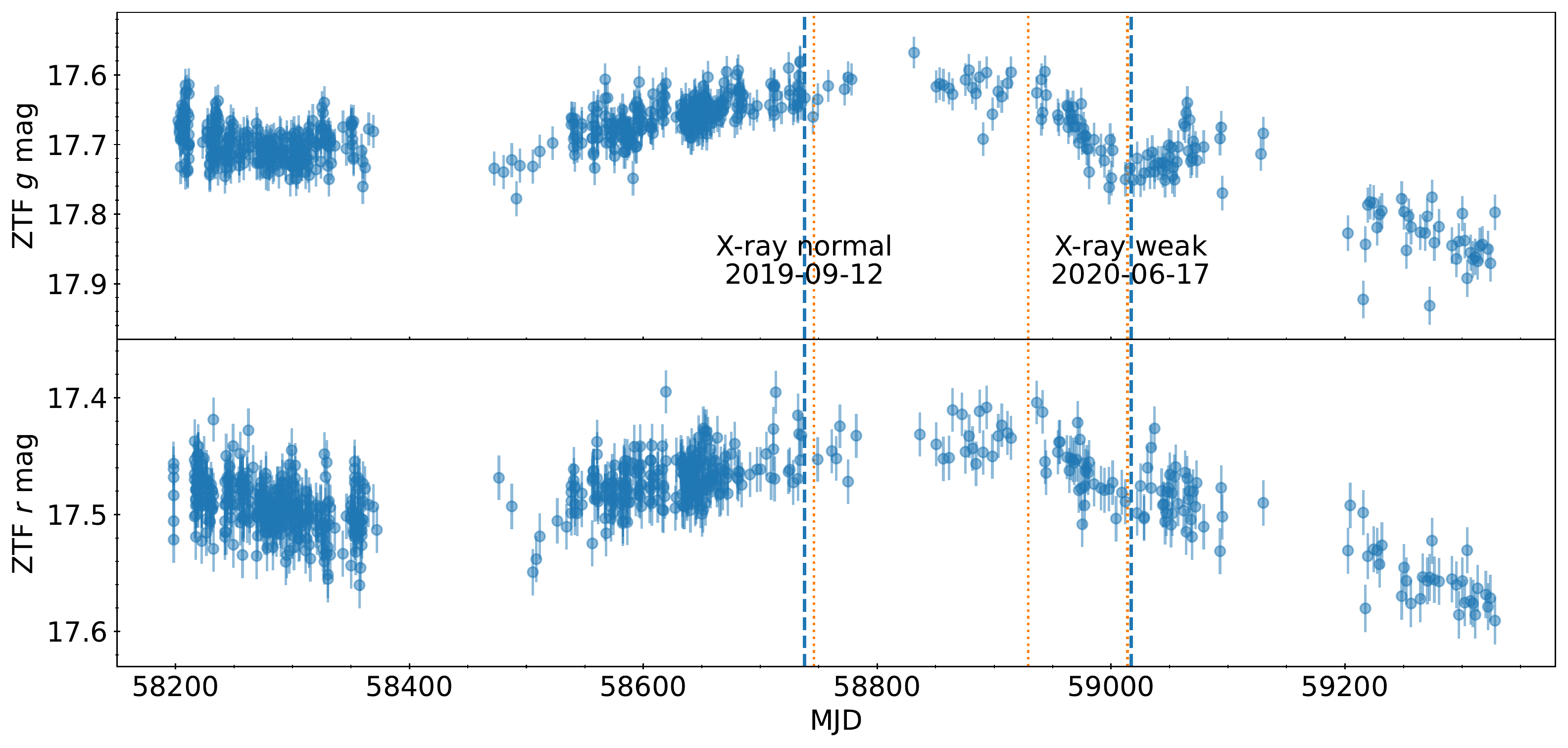}
\caption{The $g$/$r$-band lightcurves of SDSS~J1539+3954 from ZTF. 
The blue dashed lines mark the dates of the \chandra\ observations, and the orange dotted lines mark the dates of the HET observations.
Note that the observed variation in the $g$/$r$-band magnitude between the X-ray normal state and the X-ray weak state is typical among luminous quasars.
}
\label{ztflc}
\end{figure*}

Among 12 WLQs in our sample that have been observed by \chandra\ at least twice, SDSS~J1539+3954 is the only WLQ that has so far shown extreme X-ray variability.
Particularly, it has experienced two extreme X-ray state changes within 7 years (see Figure~\ref{xlc}).
We note that the 2020 \chandra\ observation of SDSS~J1539+3954 was taken with \chandra\ Director's Discretionary Time, aiming to investigate further the X-ray variability of this object. We did observe an extreme X-ray flux change again within nine months (which is $\approx 3$ months in the rest frame); specifically, the flux dropped by a factor of $\gtrsim 9$ between 2019 September and 2020 June to return this quasar to an X-ray weak level.
As stated in \citet{Ni2020}, the Hobby-Eberly Telescope (HET) observation taken contemporaneously with the 2019 September \chandra\ observation shows that the UV continuum level of this object remains generally unchanged despite the dramatic increase in the X-ray flux, and its emission lines remain weak.\footnote{While the eBOSS spectrum of SDSS~J1539+3954 taken in 2017 shows relatively higher FUV flux level (by a factor of $\approx$1.3--1.4) compared to other spectra taken at other epochs, this variation is consistent with the typical FUV variability of quasars (e.g. \citealt{Welsh2011}).}
Our new HET observation in 2020 June taken contemporaneously with the latest \chandra\ observation further confirms this (see Figure~\ref{hetspec}), considering the $\approx 20\%$ uncertainty of HET flux calibration.
The photometric data collected by the Zwicky Transient Facility (ZTF; \citealt{Bellm2019}), displayed in Figure~\ref{ztflc}, also show the variation of the $g$/$r$-band magnitude of SDSS~J1539+3954 between 2018 and 2021 (when compared to the median magnitude value in this time range) is $\lesssim 0.2/0.1$ mag, consistent with the expectations for typical quasars (e.g. see figure~3 of \citealt{MacLeod2012}).

This extreme X-ray variability of SDSS~J1539+3954 could be explained in the context of the TDO model.
We naturally expect the transition between an X-ray weak state and an X-ray normal state when there is a slight change in the
thickness of the TDO that moved across our line of sight, and the observed UV continuum/emission-line properties will not be significantly affected.
While a significant change in the global TDO structure could take up to years, if the extreme X-ray variability is caused by slight variations in the
height of the TDO, such as due to the rotation of an azimuthally asymmetric TDO, the timescale of an X-ray state transition could be small (e.g.,
weeks-to-months in the rest frame), which could explain the small timescale ($\approx 3$ months) of the X-ray state transition we recently observed for SDSS~J1539+3954.

In \citet{Ni2020}, we noted that the extreme \xray\ variability of SDSS~J1539+3954
is reminiscent of that previously found for another WLQ, PHL~1092 at $z=0.39$, which
showed a flux dimming and then re-brightening by a factor of $\approx 260$ 
during \hbox{2003--2010} (e.g. \citealt{Miniutti2012}). This similarity suggests that
weak UV emission lines may be an effective indicator for finding extreme \xray\
variability among luminous quasars, especially when considering the limited \xray\
monitoring performed thus far for WLQs. Moreover, the $z=0.18$ quasar PDS~456
has similar C~{\sc iv} properties to WLQs (e.g. \citealt{OB2005}) and shows notably
large-amplitude \xray\ variability (e.g. \citealt{Reeves2020} and references
therein), further strengthening the likely connection.\footnote{We note that the \civ\ REW of PDS~456 is variable on multi-year timescales (while the emission-line strength remains generally weak); see Fig. 2 of \citet{Hamann2018}.}
 Additional \xray\ monitoring
of WLQs is thus warranted. We also note that local Narrow-Line Seyfert~1 galaxies
(NLS1s), generally thought to have high Eddington ratios, have been proposed
to follow a physical model with broad similarities to our TDO scenario for WLQs
(e.g. \citealt{DJ2016,Hagino2016}) and that some NLS1s show extraordinary
\xray\ variability (e.g. \citealt{Boller1997, Boller2021, Leighly1999, Parker2021}).
Some of this \xray\ variability may also be caused by motions of a TDO across the
line of sight.

In addition to confirming the link between weak UV emission lines and extreme \xray\
variability to support the TDO model, X-ray monitoring of WLQs can also help reveal whether the TDO wind varies dramatically itself (e.g. via the motions of internal clumps), and is able to cause X-ray state transitions as well.
If a large fraction of WLQs exhibit extreme variability during long-term X-ray monitoring, it would suggest that their X-ray state transitions are not purely the results of slight changes in the
height of a TDO that moved across our line of sight, as we should only observe such extreme X-ray variability events among WLQs where our line of slight roughly skims the ``surface'' of the TDO according to the model.
Alternatively, if only a small fraction (though this fraction should still be considerably larger compared to that among typical quasars) of WLQs show extreme X-ray variability, it would suggest that the internal motion of the TDO is unlikely to be strong enough to lead to X-ray state transitions.
Currently, multi-epoch ($\approx 2$) \chandra\ observations have been obtained for 12 WLQs in our sample, and SDSS~J1539+3954 is the only one showing extreme X-ray variability. We will be able to obtain more information after further monitoring these WLQs.

\section{Conclusions and Future Work}

In this work, we present the observational results for a set of WLQs in a representative WLQ sample that only had limited X-ray constraints previously, and perform statistical analyses to study this representative sample. The key points are listed below:

\begin{enumerate}
\item We observed 12 X-ray undetected WLQs in the \citet{Ni2018} representative sample with deeper \chandra\ observations, and detected 7 of these WLQs. 
We also re-observed 2 WLQs in the \citet{Ni2018} sample with \chandra\ that required more secure X-ray data (see Section~\ref{s-chandra}).

\item We performed photometric analyses to estimate the X-ray fluxes of the WLQs with new \chandra\ observations (see Section~\ref{ss-xray}).
For WLQs that are still undetected, tight \xray\ flux upper limits are obtained. 
For one of the WLQs in our sample, SDSS~J1539+3954, we found that after an observed \xray\ flux rise by a factor of $\gtrsim 20$ reported in \citet{Ni2020}, it changed from an X-ray normal state back to an X-ray weak state within $\approx$ 3 months in the rest frame, with a downward variation in the \xray\ flux by a factor of $\gtrsim$ 9. 
The TDO model proposed for WLQs can explain these dramatic X-ray flux variations (see Section~\ref{ss-xvar}).

\item We updated the X-ray-to-optical properties of the \citet{Ni2018} representative WLQ sample, and compared the \daox\ distribution of this WLQ sample with that of typical quasars. We found that the \daox\ values of WLQs differ from those of typical quasars significantly, mainly because of a strong tail toward X-ray weakness.
X-ray stacking analyses show that for 5 out of 32 WLQs that are still X-ray undetected, they are X-ray weak by an average factor of $> 85$ (see Section~\ref{ss-aox}).

\item The \daox\ distribution of WLQs in our representative sample suggests that the TDO, if responsible for the \xray\ properties of WLQs, has an average global covering factor of $\sim$ 0.5. The column density of the TDO among different objects may range from $N_{\rm H}$ $\sim 10^{23-24}~{\rm cm}^{-2}$ to $N_{\rm H}$ $\gtrsim 10^{24}~{\rm cm}^{-2}$, causing different levels of absorption and Compton reflection (and/or scattering). This is also broadly supported by the non-bimodal nature of the \daox\ distribution (see Section~\ref{ss-tdo}).

\item We found that the scatter of WLQs around the \daox\ vs.\ \civ\ REW and the \daox\ vs.\ \heii~REW relations for typical quasars are significantly larger than those of typical quasars. The differences are consistent with expectations from the TDO model (see Section~\ref{ss-daoxciv}).

\item We found that after obtaining a better characterization of the X-ray-to-optical properties of the \citet{Ni2018} representative WLQ sample, the significance levels of the correlations between \daox\ and \gi\ or \feii\ REW (which are regarded as potential tracers of X-ray weakness among WLQs) are still below $3\sigma$ (see Section~\ref{ss-tracer}). 

In the future, a larger sample of WLQs with X-ray detections that are selected in an unbiased manner will help us to investigate further the potential tracers of X-ray weakness, which will ultimately enable us to probe the nature of WLQs and test the TDO model.
Long-term \xray\ monitoring of a sample of WLQs as well as X-ray spectral observations of SDSS~J1539+3954 will also enable us to examine further the TDO model, as the TDO model predicts a certain fraction of WLQs with extreme X-ray variability (this fraction is expected to be considerably larger compared to that among typical quasars) and a high apparent level of absorption in the \xray\ weak state of SDSS~J1539+3954. 
Comparisons between WLQs and other objects known to have high Eddington ratios such as NLS1s will also help to reveal the underlying physics of these objects.

\end{enumerate}

\section*{Acknowledgements}
We thank the referee for a constructive review.
We thank C. Done and A. Laor for useful discussions about WLQ models.
We thank Greg Zeimman for help reducing the HET spectra. 
QN and WNB acknowledge support from \chandra\ \xray\ Center grants GO0-21080X and DDO-21113X, NASA grant 80NSSC20K0795, and Penn State ACIS Instrument Team Contract SV4-74018 (issued by the Chandra X-ray Center, which is operated by the Smithsonian Astrophysical Observatory for and on behalf of NASA under contract NAS8-03060). 
BL acknowledges financial support from the National Natural Science
Foundation of China grant 11991053, China Manned Space Project grants
NO. CMS-CSST-2021-A05 and NO. CMS-CSST-2021-A06.

The Chandra ACIS Team Guaranteed Time Observations (GTO) utilized were selected by the ACIS Instrument Principal Investigator, Gordon P. Garmire, currently of the Huntingdon Institute for X-ray Astronomy, LLC, which is under contract to the Smithsonian Astrophysical Observatory via Contract SV2-82024.

This research has made use of data obtained from the \chandra\ Data Archive, and software provided by the \chandra\ X-ray Center in the application package CIAO.

\section*{DATA AVAILABILITY}

The data used in this investigation are available in the article.

\clearpage
\begin{landscape}
\begin{deluxetable}{lccccccccccc}
\setlength{\tabcolsep}{4pt}
\tabletypesize{\footnotesize}
\tablewidth{0pt}
 \tablecaption{New X-ray Observations and Photometric Properties of WLQs}
 \tablehead{
 \colhead{Object Name}                &
 \colhead{RA}                                &
 \colhead{Dec}                                &
 \colhead{Redshift}                        &
 \colhead{Observation}                  &
 \colhead{Observation}                  &
 \colhead{Exposure}                   &
 \colhead{Soft Band}                   &
 \colhead{Hard Band}                   &
 \colhead{Band}                   &
 \colhead{$\Gamma_{\rm eff}$} &
 \colhead{Ref.} \\
 \colhead{(J2000)}   &
 \colhead{(deg)}   &
 \colhead{(deg)}   &
 \colhead{}   &
 \colhead{ID}   &
 \colhead{Start Date}   &
 \colhead{Time (ks)}   &
 \colhead{(0.5--2~keV)}   &
 \colhead{(2--8~keV)}   &
 \colhead{Ratio}                   &
 \colhead{}                   &
 \colhead{}   \\
 \colhead{(1)}         &
 \colhead{(2)}         &
 \colhead{(3)}         &
 \colhead{(4)}         &
 \colhead{(5)}         &
 \colhead{(6)}         &
 \colhead{(7)}         &
 \colhead{(8)}         &
 \colhead{(9)}         &
  \colhead{(10)}         &
    \colhead{(11)}         &
 \colhead{(12)}        \\
 \noalign{\smallskip}\hline\noalign{\smallskip}
 \multicolumn{12}{c}{Previously X-ray undetected objects in the representative sample} 
}
 \startdata
$082508.75+115536.3$ & 126.286476    &11.926766  &    1.992   & 22527  & 2020-09-29  & 6.6 & $<4.0$ & $<2.5$ & ... & ... &     \citet{Luo2015} \\
$082722.73+032755.9$ &   126.844727   & 3.465551 &   2.031   &  22533 & 2019-09-30  & 10.1 & $3.0^{+3.1}_{-1.7}$ & $<5.9$ & $< 2.37$ &  > 0.4 &     \citet{Luo2015} \\
                                          &                         &                 &              &  22865 & 2019-10-01 & 10.0 & $2.0^{+2.8}_{-1.3}$& $<5.9$ & $< 3.89$  &  > 0.1  &     \citet{Luo2015} \\
$094533.98+100950.1$   &  146.391617 & 10.163917 & 1.672  & 22529    & 2019-10-27 & 12.0 & $4.1^{+3.3}_{-2.0}$ & $5.3^{+3.7}_{-2.4}$ &  $1.3^{+1.4}_{-0.9}$  & $0.9^{+0.8}_{-0.6}$   &  \citet{Wu2012} \\
 $095023.19+024651.7$ & 147.596664 & 2.781048 & 1.882&  22532 &  2020-01-25&   15.7  & < 5.6 &  < 5.9 & ... & ... &     \citet{Ni2018} \\
 $100517.54+331202.8$ & 151.323105 &  33.200783&  1.802 & 22535  & 2020-01-17 & 23.3 &  $13.3^{+4.9}_{-3.7}$ &   $9.4^{+4.5}_{-3.2}$ & $0.8^{+0.5}_{-0.3}$  & $1.2 ^{+0.3}_{-0.4}$ &     \citet{Luo2015} \\
 $110409.96+434507.0$ & 166.041519 & 43.751972 & 1.804& 22708  & 2020-02-10 & 15.1  & < 2.4 & $2.9^{+3.2}_{-1.8}$& > 2.80 &  < 0.3 &     \citet{Ni2018} \\
 $122311.28+124153.9$ & 185.797028 & 12.698329 & 2.068& 22709  & 2020-03-22 & 15.6  & $10.3^{+4.5}_{-3.2}$ & $8.4^{+4.4}_{-3.0}$ & $0.8^{+0.6}_{-0.4}$ & $1.2^{+0.5}_{-0.4}$ &     \citet{Ni2018} \\
 $122855.90+341436.9$ & 187.232941 & 34.243595 & 2.147& 22530 & 2020-02-26 &  14.7 & $4.1^{+3.3}_{-2.0}$ & $<5.9$ & $<1.74$ & > 0.6&     \citet{Ni2018} \\
 $134601.28+585820.2$ & 206.505341  & 58.972279 & 1.664 & 22531 & 2020-04-09  &12.4  & < 2.4 & < 4.2 & ... & ... &     \citet{Luo2015} \\
 $140710.26+241853.6$ & 211.792786 & 24.314896 & 1.668 &  22534 & 2020-08-23 &  17.5 & $7.1^{+3.9}_{-2.7}$ & $4.0^{+3.5}_{-2.1}$ & $0.6^{+0.6}_{-0.4}$  & $1.5^{+0.7}_{-0.5}$ &     \citet{Luo2015} \\
 $153913.47+395423.4$ & 234.806137  &  39.906513  & 1.930  & 23132  & 2020-06-17 & 5.0 & < 2.4 & < 2.5 & ... & ... &     \citet{Luo2015}; \citet{Ni2020}\\
 $163810.07+115103.9$ & 249.541992 & 11.851092 &1.983 & 22710 & 2020-01-04 & 13.1& < 2.4 & < 4.2 & ...  & ... &     \citet{Ni2018} \\
 \noalign{\smallskip}\hline\noalign{\smallskip}
 \multicolumn{12}{c}{Objects in the representative sample that previously lacked \chandra\ observations } \\
 \noalign{\smallskip}\hline\noalign{\smallskip}
 $       113949.39+460012.9 $& 174.955795 & 46.003604 & 1.859&   22712  & 2020-07-22 &  4.5 & $16.7^{+5.3}_{-4.1}$ &  $7.6^{+4.2}_{-2.8}$ & $0.5^{+0.3}_{-0.2}$ & $1.6^{+0.3}_{-0.4}$ &     \citet{Ni2018} \\
$       123326.03+451223.0 $& 188.358459 & 45.206402 &1.966 &  22711 &2020-08-15 &  2.7  & $25.0^{+6.2}_{-5.1}$  &  $18.7^{+5.7}_{-4.5}$ & $0.7^{+0.3}_{-0.2}$ & $1.2^{+0.3}_{-0.2}$ &     \citet{Ni2018} \\
 \enddata
 \tablecomments{
Col. (1): Object name in the J2000 coordinate format. 
Cols. (2)--(3): The SDSS position in decimal degrees.
Col. (4): Redshift adopted from \citet{Hewett2010}.
Cols. (5)--(6): The ID and start date of the \chandra\ observations.
Col. (7): Effective exposure time in the full band (0.5--8~keV) with background flares cleaned.
Cols. (8)--(9): Source counts (aperture-corrected) in the soft band (0.5--2~keV) and hard band (2--8~keV). 
If the source is undetected in a band, an upper limit of counts at a 90\% confidence level is listed.
Col. (10): Ratio of the hard-band and soft-band counts.  ``...'' indicates that the source is not detected in both bands.
Col. (11): Effective power-law photon index in the \hbox{0.5--8~keV} band. ``...'' indicates that $\Gamma_{\rm eff}$ cannot be constrained.
Col. (12): Reference paper for the object where its previous \chandra\ observations are presented. \newline
}
 \label{xraytable}
\end{deluxetable}
\end{landscape}

\begin{landscape}
\begin{deluxetable}{lccccccccccc}
\setlength{\tabcolsep}{4pt}
\tabletypesize{\footnotesize}
\tablewidth{0pt}
 \tablecaption{Stacked X-ray Photometric Properties of Previously X-ray Undetected WLQs}
 \tablehead{
 \colhead{Object Name}                &
  \colhead{Observation}                &
    \colhead{Single-epoch Count} &
 \colhead{Total Exposure}                   &
 \colhead{Stacked Soft Band}                   &
 \colhead{Stacked Hard Band}                   &
 \colhead{Band}                   &
 \colhead{$\Gamma_{\rm eff}$} &
 \colhead{Ref.} \\
 \colhead{(J2000)}   &
  \colhead{IDs}   &
    \colhead{Rate (0.5--2 keV)} &
 \colhead{Time (ks)}   &
 \colhead{(0.5--2~keV) Counts}   &
 \colhead{(2--8~keV)  Counts}   &
 \colhead{Ratio}                   &
 \colhead{}                   &
 \colhead{}   \\
 \colhead{(1)}         &
 \colhead{(2)}         &
 \colhead{(3)}         &
 \colhead{(4)}         &
 \colhead{(5)}         &
 \colhead{(6)}         &
 \colhead{(7)}         &
 \colhead{(8)}         &
 \colhead{(9)}       \\
}
 \startdata
$082508.75+115536.3$ & 14951, 22527 &  $<0.47$, $<0.60$            & 11.7 & $< 4.0$ & $< 2.5$ &  ... & ...  &   \citet{Luo2015} \\ 
$082722.73+032755.9$ & 15342, 22533, 22865 & $< 0.97$, 0.30, 0.20& 22.5 & $5.1^{+2.7}_{-2.1}$ & $3.8^{+2.5}_{-1.8}$  & $0.7^{+0.6}_{-0.5}$ & $1.3^{+0.7}_{-0.5}$ &  \citet{Luo2015} \\
$094533.98+100950.1$ & 12706, 22529 & $<1.04$, 0.45                      &12.0 & $4.1^{+2.6}_{-1.8}$ & $5.1^{+2.8}_{-2.1}$ & $1.2^{+1.0}_{-0.7}$ & $0.9^{+0.7}_{-0.4}$ &  \citet{Wu2012} \\
 $095023.19+024651.7$ & 18118, 22532& $< 0.85$, $< 0.36$               &20.5& $3.0^{+2.3}_{-1.5}$ &      < 7.6                  &  $<3.2$  &  $> 0.2$ &  \citet{Ni2018} \\
 $100517.54+331202.8$ & 15351, 22535 & $< 1.38$, 0.60                     &23.3 & $13.6^{+4.2}_{-3.5}$ & $9.1^{+3.6}_{-2.9}$ & $0.7^{+0.3}_{-0.3}$ & $1.3^{+0.4}_{-0.3}$ &   \citet{Luo2015} \\
 $110409.96+434507.0$ & 18119, 22708 & $< 0.53$, $<0.16$                &19.7 & $<2.4$                  &     < 7.6                        & ... &  ... &  \citet{Ni2018} \\
 $122311.28+124153.9$ & 18115, 22709 & $<0.89$, 0.66                       & 20.1 & $11.5^{+3.9}_{-3.2}$ & $9.1^{+3.6}_{-2.9}$ & $0.8^{+0.4}_{-0.3}$ & $1.2^{+0.4}_{-0.3}$ &  \citet{Ni2018} \\
 $122855.90+341436.9$ & 18111, 22530 & $<1.05$, 0.37                      &14.7 &  $5.0^{+2.7}_{-2.0}$ &  < 5.9                     &  < 1.4 & > 0.8 &   \citet{Ni2018} \\
 $134601.28+585820.2$ & 15336, 22531& $<1.57$, $< 0.22$               & 12.4 & < 2.4                     &     < 4.2                        & ... & ...  &  \citet{Luo2015} \\
 $140710.26+241853.6$ & 15345, 22534 & $<0.97$, 0.41                     & 20.0 &  $7.3^{+3.2}_{-2.5}$ &  $3.8^{+2.6}_{-1.8}$ &  $0.5^{+0.4}_{-0.3}$ & $1.5^{+0.7}_{-0.4}$ &  \citet{Luo2015} \\
 $153913.47+395423.4^*$ & 14948, 22528, 23132 & $<0.45$, 3.98,$<0.48$ & 10.3 & < 2.4 &  < 2.5 & ...  & ... &    \citet{Luo2015}; \citet{Ni2020}\\
 $163810.07+115103.9$ & 18116, 22710  & $<0.59$, $<0.18$               & 17.2 & < 2.4                     &     < 4.2                        & ... & ... &  \citet{Ni2018} \\
  \enddata
 \tablecomments{
Col. (1): Object name in the J2000 coordinate format. 
Col. (2): The IDs of all \chandra\ observations of this object.
Col. (3): The 0.5--2 keV \chandra\ count rate in each \chandra\ observation.
Col. (4): Total effective exposure time in the full band (0.5--8~keV) with background flares cleaned of all the available \chandra\ observations.
Cols. (5)--(6): Stacked source counts (aperture-corrected) in the soft band (0.5--2~keV) and hard band (2--8~keV) of all the available \chandra\ observations.
If the source is undetected in this band, an upper limit of counts at a 90\% confidence level is listed.
Col. (7): Ratio of the hard-band and soft-band counts.  ``...'' indicates that the source is not detected in both bands.
Col. (8): Effective power-law photon index in the \hbox{0.5--8~keV} band. ``...'' indicates that $\Gamma_{\rm eff}$ cannot be constrained.
Col. (9): Reference paper of the object where its previous \chandra\ observations are presented. 
\newline
$^*$SDSS J153913.47+395423.4 is a WLQ that has shown extreme X-ray variability, as can be seen from Col. (3). 
The X-ray stacking analyses of this object are performed only with \chandra\ observations that are consistent with an X-ray weak state.
}
 \label{stackxraytable}
\end{deluxetable}
\end{landscape}

\clearpage
\begin{landscape}
 \begin{deluxetable}{lccccccccccccc}
  \tabletypesize{\footnotesize}
  \tablewidth{0pt}
  \tablecaption{X-ray and Optical Properties of WLQs with New \chandra\ Observations}
  \tablehead{
\colhead{Object Name}                   &
\colhead{$M_{i}$}                   &
\colhead{$N_{\rm H,Gal}$}                   &
\colhead{Count Rate}                   &
\colhead{$F_{\rm X}$}                   &
\colhead{$f_{\rm 2~keV}$}                   &
\colhead{$\log L_{\rm X}$}                   &
\colhead{$f_{\rm 2500~\textup{\AA}}$}                   &
\colhead{$\log L_{\rm 2500~\textup{\AA}}$}                   &
\colhead{$\alpha_{\rm OX}$}                   &
\colhead{$\Delta\alpha_{\rm OX}(\sigma)$}                   &
\colhead{$f_{\rm weak}$}                   \\
\colhead{(J2000)}   &
\colhead{}   &
\colhead{} & 
\colhead{(0.5--2~keV)}   &
\colhead{(0.5--2~keV)}   &
\colhead{}   &
\colhead{(2--10 keV)}   &
\colhead{}   &
\colhead{}   &
\colhead{}   &
\colhead{}   &
\colhead{}   \\
\colhead{(1)}         &
\colhead{(2)}         &
\colhead{(3)}         &
\colhead{(4)}         &
\colhead{(5)}         &
\colhead{(6)}         &
\colhead{(7)}         &
\colhead{(8)}         &
\colhead{(9)}         &
\colhead{(10)}         &
\colhead{(11)}         &
\colhead{(12)}         \\
\noalign{\smallskip}\hline\noalign{\smallskip}
\multicolumn{12}{c}{X-ray undetected objects in the representative sample}
}
  \startdata
$      082508.75+115536.3$&$  -28.70$&$  3.67$&$                        <0.34$&$        <0.23$&$  <0.78$&$  <43.89$&$   4.62$&$        31.66$&$     <-2.21$&$ <-0.49(3.38)$&$ >18.4$ \\
$      082722.73+032755.9$&$  -27.62$&$  3.73$&$                          0.23$&$          0.18$&$    0.59$&$  43.93$&$   1.68$&$        31.24$&$      -2.09$&$    -0.45(2.77)$&$  14.7$ \\
$       094533.98+100950.1$&$  -27.80$&$  2.90$&$                         0.34$&$         0.23$&$     0.60$&$  43.94$&$   3.42$&$        31.40$&$      -2.05$&$    -0.53(3.46)$&$   24.4$ \\
$       095023.19+024651.7$&$  -27.48$&$  3.66$&$                         0.15$&$          0.12$&$    0.41$&$   43.75$&$   1.82$&$        31.22$&$       -2.17$&$   -0.53(3.23)$&$    23.6$ \\
$       100517.54+331202.8$&$  -26.96$&$  1.45$&$                         0.58$&$          0.46$&$    1.47$&$   44.22$&$  1.38$&$        31.07$&$      -1.87$&$    -0.30(1.75)$&$   6.0$ \\
$       110409.96+434507.0$&$   -27.29$&$  1.21$&$                       <0.12$&$       <0.10$&$    <0.33$&$   <43.68$&$  1.75$&$         31.17$&$    <-2.19$&$   <-0.57(3.43)$&$  >30.2$  \\
$       122311.28+124153.9$&$  -27.81$&$   2.52$&$                        0.57$&$         0.46$&$      1.42$&$  44.46$&$  2.05$&$         31.34$&$    -1.98$&$     -0.32(2.01)$&$  6.6$ \\
$       122855.90+341436.9$&$  -28.45$&$   1.33$&$                        0.34$&$        0.28$&$      0.96$&$  44.10$&$   3.29$&$         31.58$&$    -2.12$&$     -0.41(2.85)$&$  12.0$ \\
$       134601.28+585820.2$&$  -27.60$&$  1.63$&$                       <0.19$&$         <0.16$&$   <0.52$&$   <43.69$&$    3.22$&$       31.37$&$    <-2.22$&$   <-0.55(3.55)$&$     >27.4$ \\
$       140701.59+190417.9$&$  -26.96$&$  1.65$&$                         0.36$&$         0.32$&$     1.07$&$    43.96$&$    1.50$&$        31.04$&$     -1.97$&$    -0.37(2.15)$&$    9.2$\\
$       153913.47+395423.4^*$&$  -28.29$&$  1.70$&$                   <0.23$&$       < 0.15$&$    <0.51$&$   < 43.68$&$    3.86$&$        31.56$&$     <-2.26$&$   <-0.55(3.76)$&$  >26.7$\\
$       163810.07+115103.9$&$  -27.73$&$   4.56$&$                      <0.14$&$        <0.12$&$   <0.40$&$  <43.56$&$    2.52$&$         31.40$&$  <-2.23$&$ <-0.55(3.57)$&$  >27.0 $\\
 \noalign{\smallskip}\hline\noalign{\smallskip}
 \multicolumn{12}{c}{Objects in the representative sample that previously lacked \chandra\ observations } \\
 \noalign{\smallskip}\hline\noalign{\smallskip}
 $      113949.39+460012.9 $&$  -27.38$&$   3.85$&$                        3.80$&$         3.75$&$   13.11$&$    45.04$&$    1.92$&$        31.23$&$      -1.60$&$   0.04(0.27)$&$     0.8$ \\
$       123326.03+451223.0 $&$  -28.53$&$   4.56$&$                       9.54$&$        8.69$&$    26.66$&$     45.56$&$     4.80$&$         31.67$&$   -1.63$&$      0.10(0.68)$&$     0.6 $ \\
  \enddata
  \tablecomments{
Col. (1): Object SDSS name.
Col. (2): Absolute $i$-band magnitude.
Col. (3): The column density of Galactic neutral hydrogen \citep{Dickey1990}.
Col. (4): Soft-band (0.5--2~keV) count rate in units of 10$^{-3}$~s$^{-1}$.
Col. (5): Observed-frame 0.5--2~keV flux (corrected for Galactic absorption) in units of $10^{-14}$~\flux, assuming a power-law spectrum with $\Gamma_{\rm eff}$ derived in Section~\ref{ss-xray}.
Col. (6): Flux density at rest-frame 2~keV in units of $10^{-32}$~\mflux.
Col. (7): Logarithm of the rest-frame 2--10~keV luminosity in units of \lum, derived from $\Gamma_{\rm eff}$ and the observed-frame 0.5--2~keV flux.
Col. (8): Flux density at rest-frame 2500~\AA\ in units of $10^{-27}$~\mflux\ (from \citealt{Shen2011}).
Col. (9): Logarithm of the rest-frame 2500~\AA\ monochromic luminosity in units of \mlum.
Col. (10): Observed $\alpha_{\rm OX}$.
Col. (11): The difference between the observed $\alpha_{\rm OX}$ and the expectation from the
\hbox{$\alpha_{\rm OX}$--$L_{\rm 2500~{\textup{\AA}}}$} relation \citep{Timlin2020}. 
In the parentheses, the statistical significance of this difference is presented in units of the $\alpha_{\rm OX}$ rms scatter from Table~5 of Steffen et al. (2006).
Col. (12): The factor of X-ray weakness.
\newline
$^*$SDSS J153913.47+395423.4 is a WLQ that has shown extreme X-ray variability (\citealt{Ni2020} and Section~\ref{ss-xvar}). 
The listed X-ray properties of this WLQ are obtained from \chandra\ observations that are consistent with an X-ray weak state.}
  \label{aoxtable}
 \end{deluxetable}
\end{landscape}

\begin{table}
\caption{\heii\ measurements of WLQs}
\begin{center}
\begin{tabular}{|ccc}
\hline\hline
Object Name & MJD &  \heii\ REW \\
& & (\AA) \\
\hline
080040.77+391700.4 & 52201 & < 0.3 \\
082508.75+115536.3 & 54149 & < 0.7 \\
082722.73+032755.9 & 52642 & < 0.6 \\
084424.24+124546.5 & 53801 & 1.8 ($\pm 3.0$) \\
090312.22+070832.4 & 52674 & < 0.5 \\
094533.98+100950.1 & 52757 & < 0.7 \\
095023.19+024651.7 & 51908 & < 0.6 \\
100517.54+331202.8 & 53378 & < 1.2 \\
101209.62+324421.4 & 53442 & < 0.7 \\
101945.26+211911.0 & 53741 & 1.4 ($\pm 0.7$) \\
110409.96+434507.0 & 53053 & < 0.6 \\
113949.39+460012.9 & 53054 & < 0.7 \\
115637.02+184856.5 & 54180 & < 1.2 \\
122048.52+044047.6 & 52378 & < 0.6 \\
122311.28+124153.9 & 53120 & < 0.6 \\
122855.90+341436.9 & 53819 & < 0.4 \\
123326.03+451223.0 & 53062 & 1.7 ($\pm 0.5$)  \\
124516.46+015641.1 & 52024 & < 0.9 \\
132809.59+545452.7 & 52724 & < 0.3 \\
134601.28+585820.2 & 52425 & < 0.5 \\
140701.59+190417.9 & 54523 & < 0.4 \\
140710.26+241853.6 & 53770 & < 0.7 \\
141141.96+140233.9 & 53442 & < 0.8 \\
141730.92+073320.7 & 53499 & < 0.6 \\
150921.68+030452.7 & 52057 & 0.6 ($\pm 0.5$)  \\
153913.47+395423.4 & 53171 & < 0.3 \\
155621.31+112433.2 & 54572 & 1.6 ($\pm 0.6$) \\
161245.68+511816.9 & 52051 & < 0.7 \\
163810.07+115103.9 & 54585 & < 0.7 \\
172858.16+603512.7 & 51792 & < 1.1 \\
215954.45-002150.1 & 52173 & 0.9 ($\pm 0.3$) \\
232519.33+011147.8 & 51818 & < 0.6 \\ \hline
\end{tabular}
\end{center}
Col. (1): Object name. 
Col. (2): Modified Julian date of the SDSS observation used for measurements.
Col. (3): REWs (in units of \AA) of the \heii\ emission feature.
  \label{heiitable}
\end{table}%

\bibliography{wlq.bib}
\bibliographystyle{mnras}


\appendix

\section{\xmm\ observations of SDSS J1521+5202}
Previous studies of WLQs have stacked the counts from the X-ray weak sources, and revealed that they have hard \xray\ spectra, on average. These hard X-ray spectra suggest high levels of intrinsic X-ray absorption (\nh\ of at least $10^{23}$ cm$^{-2}$, and perhaps much greater), which
is surprising given these quasars' typical blue UV/optical continua and broad, although weak, emission lines. 
The presence of such X-ray absorption is further supported by a \chandra\ spectrum of an extremely luminous $z = 2.238$ WLQ, SDSS J1521+5202 (see Section 3.2 of \citealt{Luo2015}). 
However, the limited quality of the available X-ray spectral information remains a key barrier to further understanding.
The current 37 ks \chandra\ spectrum we obtained for J1521+5202 has 92 counts in total.
When we fit the spectrum, we are not able to distinguish between a power-law model with an intrinsic column density of \nh\ $\approx 1.3 \times 10^{23}$ cm$^{-2}$ and a Compton-reflection dominated spectral solution where the column density is  \nh~$\gg 10^{24}$~cm$^{-2}$.

Thus, we proposed an \xmm\ observation of SDSS J1521+5202 to further study whether its X-ray spectrum could be better fit with a Compton-thick reflection model or a simple absorbed power-law model.
The \xmm\ observation was split into two epochs: one observation (Observation ID: 0840440101) was conducted in July 2019, with an exposure time of 81 ks, and the other observation (Observation ID: 0840440201) was conducted in September 2019, with an exposure time of 80 ks.
After using a 3$\sigma$-clipping algorithm to remove background flaring in the MOS and removing energy ranges that are significantly affected by instrumental lines (see \citealt{Chen2018} for details), the total effective exposure times are 88.6 ks for PN, 99.5 ks for MOS1, and 137.5 ks for MOS2. 
In PN, MOS1, and MOS2, we find $73.0 \pm 18.3$, $43.1 \pm 12.4$, and $44.1 \pm 12.0$  background-subtracted counts in the \hbox{0.5--10}~keV energy range. 
The total number of source counts is $160.3 \pm 25.1$.

\begin{figure}
\centering
\includegraphics[scale=0.48]{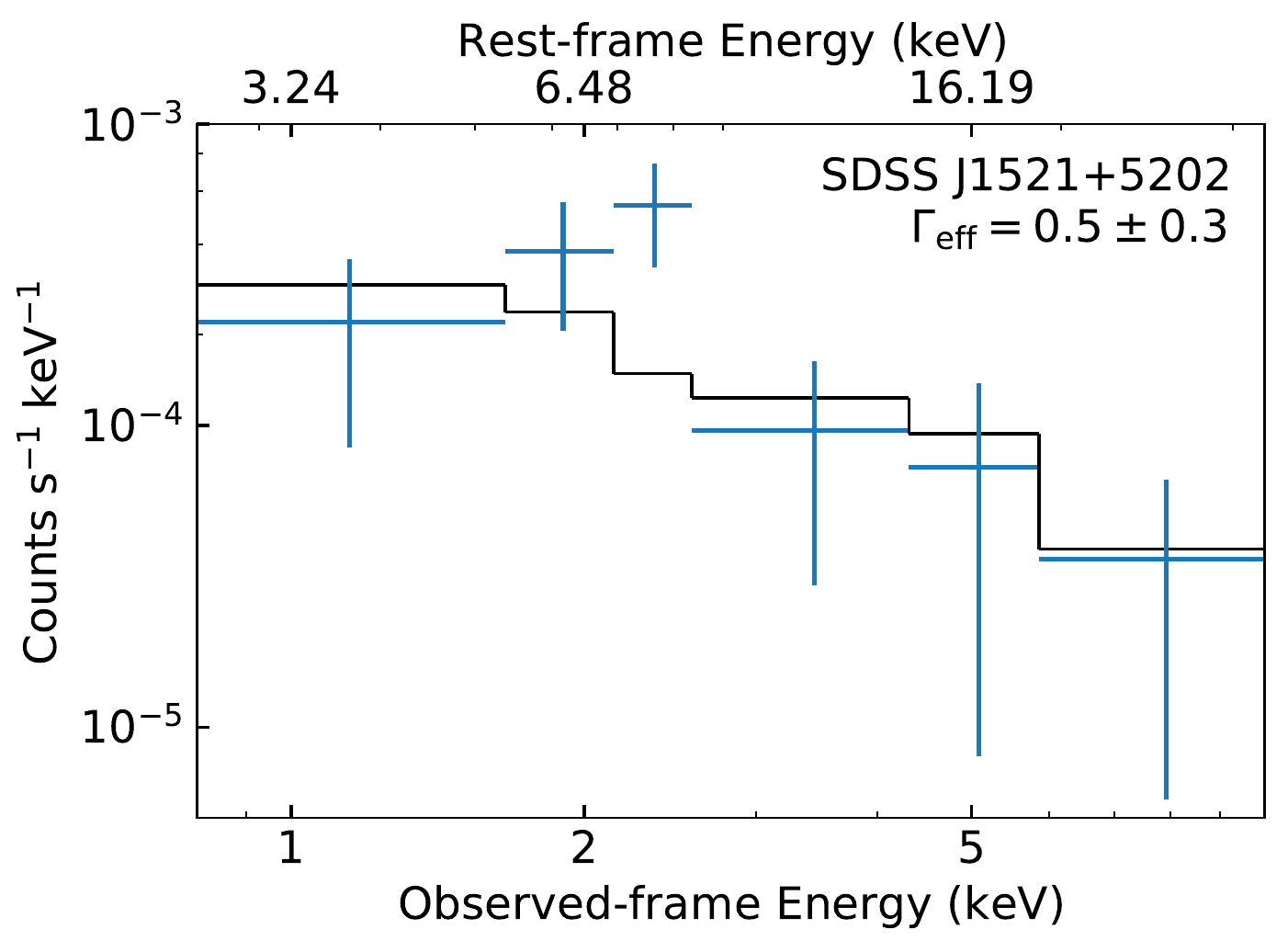}
\caption{The stacked \xmm\ EPIC spectrum of SDSS J1521+5202, shown with a folded \textit{phabs*powerlaw} model in {\sc XSPEC}.}
\label{j1521xspec}
\end{figure}

We created spectra of SDSS J1521+5202 from each instrument in each \xmm\ observation, with the XMM-Newton Science Analysis System (SAS) routine \texttt{evselect}.
Each spectrum was grouped individually, such that each bin contained a minimum of 3 counts.
Each of the individual spectra (6 in total for the 3 detectors and 2 epochs) were then combined using the SAS routine  \texttt{epicspeccombine}.
We fit the stacked spectrum with XSPEC \citep{Arnaud1996}, utilizing a power-law model modified by Galactic absorption.
The best-fit model has $\Gamma = 0.5 \pm 0.3$ (see Figure~\ref{j1521xspec}), consistent with the findings in \cite{Luo2015}.
The unabsorbed \hbox{0.5--2.0 keV} flux value of SDSS J1521+5202 at the time of \xmm\ observation is $8.7 \times 10^{-16}$ \flux, which is $\approx$ 5 times smaller compared to the X-ray flux level reported in \cite{Luo2015}.
Due to this strong flux decrease, we are not able to perform more detailed fitting to probe further the nature of the absorbing material.
As can be seen in Figure~\ref{j1521xspec}, there might also be a line emission feature at observed-frame $\approx 2.0$--2.5 keV, corresponding to rest-frame 6.4--8.0 keV. However, due to the data quality, the signal-to-noise ratio of this potential line is $< 2\sigma$, calling for further observations.
We note that similar suggestive evidence for such \xray\ line emission was found in \citet{Luo2015}.

\bsp	
\label{lastpage}
\end{document}